\documentclass[%
 reprint,
 amsmath,amssymb,
aps,
prb, 
longbibliography,
usenames,dvipsnames
]{revtex4-1}

\setcitestyle{numbers,square}
\usepackage{graphicx}
\usepackage{dcolumn}
\usepackage{bm}
\usepackage[markings, customcolors]{hf-tikz}
\usepackage{circuitikz}
\usetikzlibrary{shapes}
\usepackage{caption}
\usepackage{subcaption}
\usepackage{booktabs}
\usepackage{xcolor}
\usepackage{fancybox}
\usepackage{hyperref}

\usepackage{braket}

\definecolor{DarkGreen}{RGB}{0,75,0}

\newcommand*{\affmark}[1][*]{\textsuperscript{#1}}

\tikzset{cross/.style={cross out, draw=black, thick, minimum size=2*(#1-\pgflinewidth), inner sep=0pt, outer sep=0pt},
cross/.default={5 pt}}

\begin{document}

\preprint{APS/123-QED}


\title{Three-qubit direct dispersive parity measurement with Tunable Coupling Qubits}

\author{A. Ciani\affmark[1,3]}
\email {ciani@physik.rwth-aachen.de}
\author{D. P. DiVincenzo\affmark[1,2,3]}
 \email{d.divincenzo@fz-juelich.de}
\affiliation{
\affmark[1]Institute for Quantum Information, RWTH Aachen University,                                
  D-52056
  Aachen,                              
  Germany
}

\affiliation{
  \affmark[2]Peter Gr\"{u}nberg Institute, Theoretical Nanoelectronics,
    Forschungszentrum J\"{u}lich,
  D-52425
  J\"{u}lich,
  Germany
}

\affiliation{
\affmark[3]J\"{u}lich-Aachen Research Alliance (JARA),
    Fundamentals of Future Information Technologies,
  D-52425
  J\"{u}lich,
  Germany
}


\begin{abstract}
We consider the direct three-qubit parity measurement scheme with two measurement resonators, using circuit quantum electrodynamics to analyze its functioning for several different types of superconducting qubits. We find that for the most common, transmon-like qubit, the presence of additional qubit-state dependent coupling terms of the two resonators hinders the possibility of performing the direct parity measurement. We show how this problem can be solved by employing the Tunable Coupling Qubit (TCQ) in a particular designed configuration. In this case, we effectively engineer the original model Hamiltonian by cancelling the harmful terms. We further develop an analysis of the measurement in terms of information gains and provide some estimates of the typical parameters for optimal operation with TCQs.
\end{abstract}

\maketitle


\section{Introduction}
The circuit quantum electrodynamics (QED) architecture is a promising platform for realizing small and large scale quantum information processing \cite{blais2004, wallraff2004}. Circuit QED arose as an extension of cavity QED concepts to microwave circuits, in which the role of atoms is played by electrical degrees of freedom involving some nonlinearities, usually provided by  Josephson junctions. While in cavity QED the parameters of the atom are taken as God given, in circuit QED we have the possibility both to engineer these parameters and also to actively tune them. In order to perform quantum information processing we of course need the ability to perform high fidelity single and two-qubit gates.  But quantum measurements are equally necessary, and, to make quantum computing fault tolerant, measurements associated with quantum error correction codes are also essential.\cite{nielsenChuang, terhal2015}.\par

The present paper tackles the analysis of direct multi-qubit measurements, but a major theme of this work is the larger one of the design and analysis of couplings in mutli-qubit, multi-resonator systems.  We will see that the structure of interactions available in this system is not, at first sight, suitable for accomplishing our measurement task.  However, we will show that a broader view, in which the definition of the qubit itself is modified, in a way that is clearly experimentally feasible, greatly broadens the set of available coupling schemes that are available.  In this broadened setting, we can come very close to the ideal requirements for direct error-correction measurements. \par

The measurements associated with all the important error-correction codes, while they can be accomplished by single-qubit measurements, are more fundamentally the detection of parity properties of collections of qubits.  The traditional way of performing these so-called stabilizer measurements requires a quantum circuit involving a sequence of CNOT gates, where the code qubits are the controls and an ancilla qubit is the target. Information about the stabilizer operator to be measured is thus encoded in the state of the ancilla, which is then read out. In circuit QED this is usually done via a dispersive measurement \cite{blais2004, gambetta2008}. Although conceptually very simple, the fact that we need to run a quantum circuit in order to perform a stabilizer measurement means that the overall fidelity of the measurement would depend on the overall fidelity of a sequence of CNOT gates. The added overhead of the required ancilla qubits is another deficiency of this paradigm. For this reason the quest for alternative ways of performing stabilizer measurements has become an active area of research \cite{kerckhoff2009, lalumiere2010, tornbergJohannson2010, diVincenzoSolgun, tornbergBarzanjeh, criger2016, govia2015, govenius2015, niggGirvin, blumoff2016}.\par

This paper provides new insights into the high-fidelity implementation of multi-qubit parity measurements. Our approach begins with the scheme originally proposed in \cite{diVincenzoSolgun} and further analyzed in \cite{tornbergBarzanjeh, criger2016}.  While this approach will work for any number of qubits, we focus on the useful case \cite{bravyiSubSurfaceCode} of three qubits. The scheme is basically a direct three-qubit dispersive readout in which, by using two readout resonators, and via suitable choice of parameters, it is possible to set up conditions in which the output field depends only on the parity and not on the particular state of the three qubits. The idea is similar to the one in \cite{lalumiere2010}, which provides a solution for performing a two-qubit parity measurement with one readout resonator.\par 

In previous works \cite{diVincenzoSolgun, tornbergBarzanjeh, criger2016} it was simply assumed that a simple coupling Hamiltonian could be achieved, in which each resonator acquires a qubit-state dependent dispersive frequency shift from each qubit to which it is coupled. In this case direct three-qubit measurement is possible, and the conditions for this were identified. However, here we show that, starting with realistic (Jaynes-Cummings) couplings, the desired simple effective Hamiltonian is not straightforwardly realized, because there arise to the same order (in the Jaynes-Cummings parameter) qubit-state dependent {\em coupling} between pairs of resonators. Such a term has been previously recognised and analyzed in the literature \cite{mariantoni2008, reuther2010}, where it was referred to as the  {\em quantum switch}, because it is able to turn on and off the interaction between two cavity modes via manipulation of the state of a coupling qubit. The rotating wave approximation (RWA) would argue that these terms can be neglected when the resonators' frequencies are far detuned from each other.  Unfortunately, in the regime in which the parity measurement works the RWA is definitely inapplicable, because in that case the resonators' frequencies are constrained to be close to each other.\par

Faced with this problem, this paper considers the strategy of introducing a composite object to act an an effective qubit whose form of couplings (with the resonators and with other qubits) can be different.  In fact we have not had to search far to find such a construction: we find that the so-called Tunable Coupling Qubit (TCQ), which has been well studied since its introduction in 2011 \cite{tcq2011}, can be adopted to produce exactly the structure of couplings that we want in our application.The TCQ simply consists of two ordinary qubits (of the transmon type) with strong direct capacitive coupling between them.  As in previous work, a qubit can be defined in this multi-level system as simply the two lowest energy states.  A full analysis give here shows that the quantum switch interaction can be completely suppressed in this setting.  We hope that this work provides an example of using the remarkable flexibility of circuit QED to arrive at a purpose-built design in which desired multi-qubit functionality is achieved. \par 

The paper is organized as follows. In Sec. \ref{Sec2}, after developing the general input-output theory for two resonators coupled to a common bath of harmonic oscillators (which would be a transmission line in our case), we apply it to obtain the condition for obtaining a three-qubit parity measurement. In Sec. \ref{sec3qstransmon}, we show that these conditions cannot be matched for the case of the transmon qubit \cite{koch2007}, and also in general for a simple two-level system. The basic reason is that after obtaining the effective Hamiltonian for the system by using a Schrieffer-Wolff transformation \cite{Bravyi20112793}, the quantum switch coupling term cannot be freely chosen and is expressed as a function of the dispersive shifts of the two resonators. In this case, we cannot access the regime of parameters that allows a parity measurements with these kinds of qubits. \par 

 In Sec. \ref{sec4}, we show how this problem can be solved by using TCQs \cite{tcq2011}, which is a more flexible effective qubit than the transmon. In particular, we show how it is also possible to cancel completely the quantum switch terms, while still retaining qubit-state dependent frequency shifts on both resonators. In this way, the reduced Hamiltonian effectively realizes the original model proposed in \cite{diVincenzoSolgun}. We also identify the general condition for canceling the quantum switch terms and show how this can be intuitively understood by looking at the energy level diagram of the system. Building on this intuition, this reasoning may be applied not only to the TCQ, but also to different systems. \par 
 
In Sec. \ref{sec5}, using Bayesian inference, and expanding the discussion in \cite{clerkNoise} for the single qubit case, we show how to rigorously define information gain and rate of information gain of the parity of the set of qubits, and provide some estimates of the achievable parameters attainable using TCQs as qubits. We finally draw the conclusions in Sec. \ref{Sec6}.     
\section{DERIVATION OF THE PARITY CONDITION}
\label{Sec2}
In this section, we start by briefly reviewing the input-output theory for a system of two resonators, or in general two bosonic modes, coupled to the same transmission line. Afterwards, we apply this theory to a system of three qubits coupled dispersively to two resonators, in order to obtain the condition that must be fullfilled for measuring only the parity, and not the particular state, of the string of qubits.
\subsection{Two-Resonators Input-Output Theory}
In this subsection we closely follow standard references about input-output theory \cite{gardinerZollerNoise, wallsMilburn}. We consider a system in which there are two resonators coupled to the same transmission line, which is modelled as a harmonic oscillator bath. The Hamiltonian takes the following form
\begin{equation}
H= H_{sys}+H_B+H_{int}
\end{equation}
with the bath Hamiltonian $H_B$ and the interaction Hamiltonian $H_{int}$ defined as
\begin{subequations}
\begin{equation}
H_B= \int_{-\infty}^{+\infty} d \omega \omega b^{\dagger}(\omega) b(\omega), 
\end{equation}
\begin{equation}
H_{int}=\sum_{j=1}^2 i \int_{-\infty}^{+\infty} d \omega \kappa_j (\omega) \bigl[b^{\dagger}(\omega) a_j-b(\omega) a_j^{\dagger} \bigr],
\end{equation}
\end{subequations}
where, as in the references, we have made the rotating wave approximation (RWA), neglecting terms $a_j^{\dagger} b^{\dagger}(\omega)$ and $a_j b(\omega)$, and we have taken the lower limit of integration to $-\infty$ instead of $0$. In addition, the bath operators satisfy the continuous bosonic commutation relation $[b(\omega), b^{\dagger}(\omega \sp{\prime})]=\delta(\omega-\omega \sp{\prime})$, while for the system's annihilation and creation operators $[a_i, a_j^{\dagger}]= \delta_{ij}$. The system Hamiltonian $H_{sys}$ is kept completely generic for now; it can be the simple Hamiltonian of two harmonic oscillators, or something more complicated in which the bosonic modes are connected also to other systems, as we will consider in the next subsection. The important thing is that just the two bosonic modes are coupled directly to the common bath. \par 
The Heisenberg equations of motion for bath and system annihilation operators read
\begin{subequations}
\begin{equation}
\label{HeisEqbomega}
\frac{d  b(\omega)}{d t}= -i \omega b(\omega) +\sum_{j=1}^2 \kappa_j(\omega) a_j.
\end{equation}
\begin{equation}
\label{HeisEqa1}
\frac{d  a_1}{dt} = -i \bigl[a_1, H_{sys}\bigr] -\int_{-\infty}^{+\infty} d \omega \kappa_1(\omega) b(\omega),
\end{equation}
\begin{equation}
\frac{d  a_2}{dt} = -i \bigl[a_2, H_{sys}\bigr] -\int_{-\infty}^{+\infty} d \omega \kappa_2(\omega) b(\omega),
\end{equation}
\end{subequations}
Viewing the system annihilation operators in  Eq. \ref{HeisEqbomega} as forcing terms, and setting the initial condition at time $t_0 < t$ we obtain the formal solution
\begin{multline}
b(\omega;t)= e^{-i \omega(t-t_0)} b(\omega; t_0) \\ +\sum_{j=1}^2 \kappa_j (\omega) \int_{t_0}^t dt \sp{\prime} e^{-i \omega(t-t \sp{\prime})} a_j(t \sp{\prime}),
\end{multline} 
and inserting this result into Eq. \ref{HeisEqa1} we get
\begin{multline}
\label{eqa1NM}
\frac{d  a_1}{dt} = -i \bigl[a_1, H_{sys}\bigr]-\int_{-\infty}^{+\infty} d \omega e^{-i \omega(t-t_0)}\kappa_1(\omega) b(\omega; t_0) \\
+\sum_{j=1}^2 \int_{-\infty}^{+\infty} d \omega \kappa_1(\omega) \kappa_j(\omega) \int_{t_0}^t d t\sp{\prime} e^{-i \omega(t-t \sp{\prime})} a_j(t \sp{\prime})
\end{multline}
At this point, we are in a position to make the so-called \emph{first Markov approximation}, which consists in assuming that the coupling coefficients $\kappa_j(\omega)$ vary only slowly with frequency. In a field interpretation of the bath, this approximation is equivalent to the locality of the interaction between the system and the field \cite{gardinerZollerNoise}. Hence, we set 
\begin{equation}
\kappa_1(\omega)= \sqrt{\frac{\kappa_1}{2 \pi}} \quad, \quad \kappa_2(\omega)= \sqrt{\frac{\kappa_2}{2 \pi}},
\end{equation}
and within this approximation Eq. \ref{eqa1NM} becomes
\begin{equation}\
\label{eqa1in}
\frac{d  a_1}{dt} = -i \bigl[a_1, H_{sys}\bigr] -\frac{\kappa_1}{2} a_1 -\frac{\sqrt{\kappa_1 \kappa_2}}{2} a_2 -\sqrt{\kappa_1}b_{in},
\end{equation}
where we have defined the \emph{input field} 
\begin{equation}
b_{in}= \frac{1}{\sqrt{2 \pi}} \int_{-\infty}^{+\infty} d \omega e^{-i \omega (t-t_0)} b(\omega;t_0),
\end{equation}
and used the properties
\begin{subequations}
\begin{equation}
\int_{-\infty}^{+\infty} d \omega e^{-i \omega (t-t\sp{\prime})}=2 \pi \delta(t-t \sp{\prime}),
\end{equation}
\begin{equation}
\int_{t_0}^t d t \sp{\prime} f(t \sp{\prime}) \delta(t-t\sp{\prime})= \frac{f(t)}{2}.
\end{equation}
\end{subequations}
The second property may look ambiguous since the singular point of the delta function is at one extremum of the integral. The reason why it holds is that the delta function introduced here must always be defined as the limit of a sequence of even functions in time \footnote{See Subsec. 3.1.1 in Ref. \cite{gardinerZollerNoise}: the noise kernel function $f(t)$ there defined is an even function of time, and the delta originates as the limit of this function.}.   \par
Proceeding analogously with the Heisenberg equation of motion involving the derivative of $a_2$, we also get
\begin{equation}
\label{eqa2in}
\frac{d  a_2}{dt} = -i \bigl[a_2, H_{sys}\bigr]-\frac{\sqrt{\kappa_1 \kappa_2}}{2} a_1 -\frac{\kappa_2}{2} a_2  -\sqrt{\kappa_2}b_{in}.
\end{equation}
It is worth pointing out that the fact that the two bosonic modes are interacting with the same bath manifests itself in the bath induced interaction between the modes, \i.e., the terms with coefficient $\sqrt{\kappa_1 \kappa_2}/2$. This implies a correlated emission of the resonators into the bath, which can give rise to the phenomenon of resonator superradiance, in close analogy to the standard atomic superradiance \cite{petruccione}. However, we will not deal with any superradiance phenomena in this article.  \par   
If instead of the condition in the past at $t_0< t$, we specified  the future condition at time $t_1 > t$, the formal solution of Eq. \ref{HeisEqbomega} would have read
\begin{multline}
b(\omega; t)= e^{-i \omega(t-t_1)} b(\omega; t_1) \\ -\sum_{j=1}^2 \kappa_j (\omega) \int_{t}^{t_1} dt \sp{\prime} e^{-i \omega(t-t \sp{\prime})} a_j(t \sp{\prime}).
\end{multline}
Using this solution, and repeating the same calculations as before, gives the following coupled equations for $a_1$ and $a_2$
\begin{subequations}
\label{eqaout}
\begin{equation}
\frac{d  a_1}{dt} = -i \bigl[a_1, H_{sys}\bigr] +\frac{\kappa_1}{2} a_1 +\frac{\sqrt{\kappa_1 \kappa_2}}{2} a_2 -\sqrt{\kappa_1}b_{out},
\end{equation}
\begin{equation}
\frac{d  a_2}{dt} = -i \bigl[a_2, H_{sys}\bigr]-\frac{\sqrt{\kappa_1 \kappa_2}}{2} a_1 -\frac{\kappa_2}{2} a_2  -\sqrt{\kappa_2}b_{out},
\end{equation}
\end{subequations}
where we defined the \emph{output field} as 
\begin{equation}
b_{out}(t) = \frac{1}{\sqrt{2 \pi}} \int_{-\infty}^{+\infty} d \omega e^{-i \omega (t-t_1)} b(\omega;t_1).
\end{equation}
Comparing Eqs. \ref{HeisEqa1} and \ref{eqa2in} with Eqs. \ref{eqaout}, we easily obtain the \emph{input-output relation} for this system:
\begin{equation}
\label{inputOutputRel}
b_{out}(t)= b_{in}(t)+\sqrt{\kappa_1} a_1(t)+\sqrt{\kappa_2} a_2(t).
\end{equation} 
In the following, we will use this relation in order to obtain the functional form of the output field for the system we want to study. \par 
\subsection{Parity Condition}
We begin by introducing the generic Hamiltonian that we are going to treat in this article, \i.e., the dispersive Hamiltonian of a system of three qubits coupled to two resonators. At the beginning we will keep the discussion completely general, neither specifying how this Hamiltonian can be obtained, nor assuming any constraints on the value of the parameters we are going to introduce. These specific features will be treated extensively in the following sections. \par
The model Hamiltonian reads 
\begin{multline}
\label{modelH}
H= \sum_{i=1}^3 \frac{\Omega_i}{2} \sigma_i^{z} +\biggl(\omega_1+ \chi_1 \sum_{i=1}^3 \sigma_{i}^z \biggr) a_1^{\dagger} a_1 \\
+\biggl(\omega_2+ \chi_2 \sum_{i=1}^3 \sigma_{i}^z \biggr) a_2^{\dagger} a_2 + \chi_{12}\sum_{i=1}^3 \sigma_{i}^z \bigl(a_1 a_2^{\dagger}+a_1^{\dagger} a_2 \bigr).
\end{multline}
We see that in addition to the qubit-state dependent dispersive shifts $\chi_1$ and $\chi_2$ of the resonators' frequencies, we also include a qubit-state dependent coupling of the two resonators with parameter $\chi_{12}$. Such term comes up in the dispersive regime of the multi-mode Jaynes Cummings model, and in the literature it has been proposed to use them as a quantum switch, able to turn on and off the interaction between two cavity modes \cite{mariantoni2008, reuther2010}. For us this term will play a major role in our parity measurement setup. In particular, we cannot neglect it assuming that the resonators are far away in frequency and thus employing a RWA. In fact, the measurement scheme requires the resonators' frequencies to be quite close to each other, invalidating the RWA. We also notice that in our model Hamiltonian Eq. \ref{modelH}, we require equal dispersive shifts and equal quantum switch parameters for all qubits. This ensures that the evolution of the resonators' field amplitudes, and accordingly also of the output field, depends only on the Hamming weight $h_w$ (the number of ones of the three qubits), and not on the particular state. Finally, we point out that in Eq. \ref{modelH} we neglected the qubit-qubit interaction mediated by the resonators, which usually comes up in the dispersive regime of cavity QED, assuming that the qubits' frequencies are far away from each other, and invoking a RWA. \par 
The Hamiltonian Eq. \ref{modelH} implies that the evolution of the resonators' annihilation operators is dependent on the Hamming weight of the string of qubits. In particular, we have from Eqs. \ref{eqa1in}, \ref{eqa2in} 
\begin{subequations}
\label{eqahw}
\begin{multline}
\frac{d a_{1, h_w}}{d t} = -i [\omega_1+\chi_1(3-2 h_w)] a_{1, h_w}-\chi_{12}(3-2 h_w) a_{2, h_w} \\ -\frac{\kappa_1}{2} a_{1, h_w}-\frac{\sqrt{\kappa_1 \kappa_2}}{2} a_{2, h_w} -\sqrt{\kappa_1} b_{in},
\end{multline}
\begin{multline}
\frac{d a_{2, h_w}}{d t} = -i [\omega_2+\chi_2(3-2 h_w)] a_{2, h_w}-\chi_{12}(3-2 h_w) a_{1, h_w} \\ -\frac{\kappa_2}{2} a_{2, h_w}-\frac{\sqrt{\kappa_1 \kappa_2}}{2} a_{1, h_w}, -\sqrt{\kappa_2} b_{in}
\end{multline}
\end{subequations}   
with $h_{w}=\{0,1,2,3\}$. In order to achieve a parity measurement, we need to ensure that at the steady state the output field, or equivalently the reflection coefficient, depends only on the parity of the register, and not on the particular Hamming weight. Let us thus obtain the generic expression for the reflection coefficient, by Fourier transforming Eqs. \ref{eqahw}. In particular, we define the Fourier transform of a generic operator in the Heisenberg picture $c(t)$ as
\begin{equation}
c[\omega]= \frac{1}{\sqrt{2 \pi}} \int_{-\infty}^{+\infty} dt e^{i \omega t} c(t).
\end{equation} 
This gives
\begin{equation}
\begin{bmatrix}
a_{1,h_w}[\omega] \\
a_{2, h_w}[\omega]
\end{bmatrix}
= A^{-1} \begin{bmatrix}
\sqrt{\kappa_1} \\
\sqrt{\kappa_2}
\end{bmatrix}
b_{in}[\omega],
\vspace{4 mm}
\end{equation}
where we defined the matrix
\begin{footnotesize}
\begin{equation}
A=
\begin{bmatrix}
i \{\Delta_{d1}- \chi_1 (3-2 h_w)\}-\frac{\kappa_1}{2} & -i \chi_{12} (3-2 h_w) -\frac{\sqrt{\kappa_1 \kappa_2}}{2} \\ -i \chi_{12} (3-2 h_w) -\frac{\sqrt{\kappa_1 \kappa_2}}{2} & i \{\Delta_{d2}- \chi_2 (3-2 h_w\}-\frac{\kappa_2}{2}
\end{bmatrix},
\end{equation}
\end{footnotesize}
with
\begin{equation}
\Delta_{di} = \omega-\omega_{i} \quad, \quad i={1,2},
\end{equation} 
is the detuning between the frequency of resonator $i$ and the drive frequency. Solving for $a_{1, h_w}[\omega]$ and $a_{2, h_w}[\omega]$, and using the input-output relation Eq. \ref{inputOutputRel}, we can write the expression for the Fourier transformed output field operator 
\begin{equation}
b_{out}[\omega]= r(\omega; h_w) b_{in}[\omega],
\end{equation}
with the Hamming weight dependent reflection coefficient given by
{\footnotesize
\begin{widetext}
\begin{multline}
r(\omega; h_w)= 1 \\
- \frac{2 (\Delta_{d1} \kappa_2 +\Delta_{d2} \kappa_1+ (3-2 h_w)(\kappa_1 \chi_1 +\kappa_2 \chi_2-2 \sqrt{\kappa_1 \kappa_2} \chi_{12})}{\Delta_{d1} \kappa_2 + \Delta_{d2} \kappa_1 +(3-2 h_w)(\kappa_1 \chi_1 +\kappa_2 \chi_2-2 \sqrt{\kappa_1 \kappa_2} \chi_{12})+2 i [\Delta_{d1} \Delta_{d2} +(3-2 h_w)(\Delta_{d1} \chi_1+\Delta_{d2} \chi_2+(3-2 h_w)^2(\chi_1 \chi_2-\chi_{12}^2))]}.
\end{multline}
\end{widetext}
}
Notice that this reflection coefficient can be written for every Hamming weight as $r(\omega; h_w)=(i b -a)/(ib+a)$ with $a, b \in \mathbb{R}$, satisfying the condition $\lvert r(\omega; h_w) \rvert=1$, as one expects since there are no lossy elements in the system. \par 
At this point we look for specific values of the detunings $\Delta_{d1}$ and $\Delta_{d2}$ such that the reflection coefficient depends only on the parity, and it also has different values
between even and odd parity. Namely, we would like $r(\omega; h_w=0)=r(\omega; h_w=2)=r_{even}$, $r(\omega; h_w=1)=r(\omega; h_w=3)=r_{odd}$ and $r_{even} \neq r_{odd}$. In particular, this last condition ensures that we can distinguish between states of different parity. These conditions are satisfied if
\begin{subequations}
\label{parityCondition}
\begin{equation}
\Delta_{d1}= \omega-\omega_1= \pm \sqrt{3} \sqrt{\frac{\kappa_1}{\kappa_2}} \sqrt{\chi_1 \chi_2-\chi_{12}^2}, 
\end{equation}
\smallskip
\begin{equation}
\Delta_{d2}= \omega-\omega_2= \mp \sqrt{3} \sqrt{\frac{\kappa_2}{\kappa_1}} \sqrt{\chi_1 \chi_2-\chi_{12}^2}. 
\end{equation}
\end{subequations}
Thus, by appropriately selecting the drive frequency $\omega$ and the resonators' frequencies $\omega_1$ and $\omega_2$, the reflection coefficient would depend only on the parity of the state of the qubits. Notice that this means that the output field, and consequently both quadratures at the steady state, depend only on the parity. This is indeed important, since one may satisfy the parity condition for one quadrature, and measure it via homodyne detection, but the conjugate quadrature may still show a general Hamming weight dependence, which is then information that gets lost into the environment causing additional intraparity dephasing \cite{tornbergBarzanjeh}.  \par 
If we set $\chi_1=\chi_2=\chi$ and $\chi_{12}=0$, the parity condition Eqs. \ref{parityCondition} reduces to the one reported in \cite{tornbergBarzanjeh, criger2016}. However, we see the important role that the quantum switch coefficient $\chi_{12}$ plays in these equations. In particular, if $\chi_{12}^2$ is larger than the product $\chi_1 \chi_2$, we cannot match the parity condition since this would require complex detunings, which is clearly absurd. As we will show this is indeed the case if we try to obtain the effective Hamiltonian Eq. \ref{modelH} using simple two-level system and also transmon qubits, in which also the second excited level must be taken into account in order to get the correct dispersive shifts \cite{koch2007}.

\section{Quantum Switch term for a Transmon}
\label{sec3qstransmon}
We start by deriving the functional form of the $\chi_{12}$ for the case of the transmon qubit. In this derivation, we will consider only one transmon linearly coupled to two resonators. Approximating directly the transmon as a Duffing oscillator with Hamiltonian \cite{koch2007}
\begin{equation}
\label{Htransmon}
H_{duff}= \omega_t b^{\dagger} b+ \frac{\delta}{2} b^{\dagger} b^{\dagger} b b,
\end{equation}
where $b^{\dagger}$ and $b$ are creation and annihilation operators for the transmon satisfying the commutation relation $[b, b^{\dagger}]=1$. In addition, the frequency $\omega_t$ and the anharmonicity $\delta$ in Eq. \ref{Htransmon} are a function the Josepshon energy $E_J$ and charging energy $E_C$. In particular, $\omega_t= \sqrt{8 E_C E_J}-E_C$ and $\delta=-E_C$. Coupling the transmon linearly to two resonators we obtain the following total Hamiltonian of the system
\begin{multline}
\label{HtransmonRes}
H= H_{duff} + \sum_{i=1}^2 \omega_i a_{i}^{\dagger} a_{i} + g_i \bigl(a_i b^{\dagger}+ \mathrm{H.c.}\bigr),
\end{multline} 
with the $g_i$ the linear coupling constant between the resonators and the transmon. We leave the value of these parameters completely generic, but real, without loss of generality. Notice however that in Eq. \ref{HtransmonRes} we are invoking immediately a RWA neglecting terms $a_i b$ and $a_i^{\dagger} b^{\dagger}$, which would be present in the general case. \par
We are interested in obtaining an effective dispersive Hamiltonian that describes accurately the system when only the ground state $\ket{g}$ and the first excited state $\ket{e}$ of the transmon are populated. However, as shown in \cite{koch2007} for the single resonator case, in order to get the correct result it is necessary to consider also the second excited level $\ket{f}$ of the transmon before carrying out the dispersive transformation. The basic reason is that the coupling part of the Hamiltonian Eq. \ref{HtransmonRes} is able to cause transitions  $\ket{g}\leftrightarrow \ket{e}$ and $\ket{e} \leftrightarrow \ket{f}$, so that, when adiabatically eliminated both transitions would contribute to the energy shift of state $\ket{e}$ . The other levels do not contribute, since the transition diagram for a transmon coupled linearly to a resonator has a ladder like structure, as immediately seen from Eq. \ref{HtransmonRes}. \par 
Hence, we project the Hamiltonian \eqref{HtransmonRes} on the subspace spanned by the first three levels of the transmon obtaining
\begin{multline}
H= \Omega_e \ket{e}\bra{e} + \Omega_f \ket{f} \bra{f}+ \sum_{i=1}^2 \omega_i a_{i}^{\dagger} a_{i} + \\
g_i \bigl( \ket{e}\bra{g} a_i +\mathrm{H.c.} \bigr) +\sqrt{2} g_i \bigl(\ket{f}\bra{e} a_i+\mathrm{H.c.}\bigr),
\end{multline}
with $\Omega_e=\omega_t$ and $\Omega_f=2 \omega_t+\delta$. \par
Assuming that both interactions of the transmon with the resonators are in the dispersive regime, which mathematically means $\lvert g_i/(\Omega_e-\omega_i) \rvert =\lvert g_i/\Delta_i \rvert  \ \ll 1$, $\lvert \sqrt{2}g_i/[(\Omega_f-\Omega_e)-\omega_i] \rvert =\lvert \sqrt{2}g_i/(\Delta_i+\delta) \rvert \ll 1$, the generator $S$ of the first order Schrieffer-Wolff unitary transformation $D=\exp \bigl[S-S^{\dagger}\bigr]$ that removes the interaction between the transmon and the resonators is found to be \cite{Bravyi20112793, Winkler} 
\begin{equation}
S= \sum_{i=1}^2 \frac{g_i}{\Delta_i} \ket{e}\bra{g} a_i +  \frac{\sqrt{2}g_i}{\Delta_i+\delta} \ket{f}\bra{e} a_i.
\end{equation} 
Applying this transformation to the Hamiltonian Eq. \ref{HtransmonRes} and retaining only terms up to order $(g_i/\Delta_i)^2$ and $(g_i/(\Delta_i+\delta))^2$, we obtain the following effective Hamiltonian
\begin{widetext}
\label{dispHtransmon}
\begin{multline}
H_{eff}= D H D^{\dagger}= \biggl(\Omega_e+ \sum_{i=1}^2 \frac{g_i^2}{\Delta_i} \biggr) \ket{e}\bra{e} + \biggl(\Omega_f+ \sum_{i=1}^2 2\frac{g_i^2}{\Delta_i+\delta} \biggr) \ket{f}\bra{f}+ \\ \sum_{i=1}^2 \biggl \{ \biggl (\omega_i+\frac{g_i^2}{\Delta_i}(\ket{e}\bra{e}-\ket{g}\bra{g})+2 \frac{g_i^2}{\Delta_i+\delta}(\ket{f}\bra{f}-\ket{e}\bra{e})\biggr)a_i^{\dagger} a_i+\frac{1}{\sqrt{2}} \biggl(\frac{g_i^2}{\Delta_i+\delta}-\frac{g_i^2}{\Delta_i} \biggr ) \bigl (\ket{f}\bra{g}a_i a_i+\mathrm{H.c.}\bigr)+ \\
\frac{g_1 g_2}{2} \biggl [2 \biggl(\frac{1}{\Delta_1+\delta}+\frac{1}{\Delta_2+\delta} \biggr)(\ket{f}\bra{f}-\ket{e}\bra{e}) +\biggl(\frac{1}{\Delta_1}+\frac{1}{\Delta_2} \biggr)(\ket{e}\bra{e}-\ket{g}\bra{g}) \biggr] \bigl(a_1 a_2^{\dagger}+\mathrm{H.c.} \bigr)+\\
\frac{g_1 g_2}{\sqrt{2}} \biggl(\frac{1}{\Delta_1+\delta}+\frac{1}{\Delta_2+\delta}-\frac{1}{\Delta_1}-\frac{1}{\Delta_2} \biggr) \bigl(\ket{f}\bra{g} a_1 a_2^{\dagger}+\mathrm{H.c.} \bigr).
\end{multline}
\end{widetext}
The two-photon transitions terms $\ket{f}\bra{g} a_i a_j$, which are present also in the dispersive transformation for the transmon coupled to a single resonator, can safely be neglected, since they are in turn far off-resonant. Finally, projecting onto the subspace spanned by $\ket{g}$ and $\ket{e}$ and introducing the Pauli operators $\ket{e}\bra{e}=(1+\sigma^z)/2$, $\ket{g}\bra{g}=(1-\sigma^z)/2$, we obtain the effective dispersive Hamiltonian for the first two levels of the transmon
\begin{multline}
\label{HteffTLS}
H_d = \frac{\bar{\Omega}}{2} \sigma^z + \sum_{i=1}^2 \biggl(\bar{\omega}_i+\chi_i \sigma^z \biggr)a_i^{\dagger}a_i + \\ \bigl(\bar{\chi}_{12} + \chi_{12} \sigma^{z} \bigr) \bigl(a_1^{\dagger}a_2+ \mathrm{H.c.} \bigr),
\end{multline}
where the effective qubit frequency is given by $\bar{\Omega}= \Omega_e+\sum_{i=1}^2 g_i^2/\Delta_i$, while the effective resonator frequencies are $\bar{\omega}_i= \omega_i-g_i^2/(\Delta_i+\delta)$. In addition, we defined the following parameters in Eq. \ref{HteffTLS}
\begin{subequations}
\label{chitransmon}
\begin{equation}
\chi_i =  \frac{g_i^2}{\Delta_i}-\frac{g_i^2}{\Delta_i+\delta},
\end{equation}
\begin{equation}
\chi_{12}= \frac{1}{2} \biggl(\frac{g_2}{g_1} \chi_1+\frac{g_1}{g_2} \chi_2 \biggr),
\end{equation}
\begin{equation}
\bar{\chi}_{12}= - \frac{g_1 g_2}{2} \biggl (\frac{1}{\Delta_1+\delta}+\frac{1}{\Delta_2+\delta} \biggr),
\end{equation}
\end{subequations}
$i= \{1,2\}$. Notice that in the limit of anharmonicity that goes to infinity, we recover the parameters that we would obtain in case we started directly with a two-level system in place of a transmon. In this limit, we get $\bar{\omega}_i \rightarrow \omega_i$ and $\bar{\chi}_{12} \rightarrow 0$. Thus, the presence of a qubit-state independent frequency shift of the resonators and coupling of the resonators is a consequence of the finite anharmonicity of the transmon. \par 
Compared to the model Hamiltonian Eq. \ref{modelH} we treated in Sec. \ref{Sec2}, we see that each transmon will cause a modification of the resonators' frequencies, leading to two new effective frequencies of the resonators. Moreover, we also get a qubit-state independent coupling of the resonators. Including, this term in the input-output theory developed in Sec.\ref{Sec2}, would give actually a modified parity condition compared to Eq. \ref{parityCondition}, which however still relies on the assumption $\chi_{12}^2 > \chi_1 \chi_2$. As one can readily check from Eqs. \ref{chitransmon}, this condition is never satisfied for a transmon and consequently it cannot be employed directly for the dispersive three-qubit parity measurement described in Sec. \ref{Sec2}. The main problem is basically that the $\chi_{12}$ parameter is not a free parameter and it is connected to the values of $\chi_1$ and $\chi_2$. In the next section we will show how this problem can be solved, by using a system that is closely related to the transmon, but allows a more flexible tuning of the parameters.   

\section{Quantum Switch term for a TCQ}
\label{sec4}
In this section we show how it is possible to tackle the problem of the quantum switch term that was raised in the previous Sec. \ref{sec3qstransmon} using a so-called Tunable Coupling Qubit (TCQ) in a particular configuration. We start by briefly reviewing the TCQ and then discuss its potential application for the implementation of the direct three-qubit parity measurement. In particular, we will show how the harmful quantum switch term can be set ideally to zero, while still retaining a qubit-state dependent dispersive shift on each resonator. In addition, we will also explain the general idea for the elimination of the quantum switch term, so that the same reasoning will also be potentially applicable to different systems and not only a TCQ. 
\subsection{Review of the TCQ}
\begin{figure}
\centering
   \begin{circuitikz}[american voltages, scale=0.6, every node/.style={transform shape}]
      \draw[Blue, ultra thick] (0,0)
      to[V,v, color=Blue] (0,3) 
      to [C, color=Blue] (3,3)
      to [short] (3,2.5)
      to [short] (2,2.5)
      to[C, color=Blue] (2,0.5) 
      to [short] (3,0.5)
      to [short] (3,0)
      to[short] (0,0);
      \draw[Blue, ultra thick] (3, 2.5)
      to [short] (4,2.5)
      to [short] (4, 0.5)
      to [short] (3,0.5);
      \draw[Blue, ultra thick] (4, 1.5) node[cross=8,color=Blue, ultra thick,rotate=0] {};
\node at  (4.7, 1.5) {\large{$\hm{E_{J+}$}}};
\node at  (2.9, 1.5) {\large{$\hm{C_{J+}$}}};
\node at  (5, 3.7) {\large{$\hm{C_{I}}$}};
\node at  (8.5, 3.7) {\large{$\hm{C_{g-}}$}};
\node at  (1.5, 3.7) {\large{$\hm{C_{g+}}$}};
\node at  (-1, 1.5) {\large{$\hm{V_{g+}}$}};
\node at  (11, 1.5) {\large{$\hm{V_{g-}}$}};
      \draw[Brown, ultra thick] (3,3)
      to [C, color=Brown] (7,3);
      \draw[Red, ultra thick] (7,3)
      to [short] (7,2.5)
      to [short] (6, 2.5)
      to [C, color=Red] (6, 0.5)
      to [short] (7,0.5)
      to [short] (7,0)
      to [short] (5,0);
      \draw[Blue, ultra thick] (5,0)
      to [short] (3,0);
      \draw[Red, ultra thick] (7, 0.5)
      to [short] (8,0.5)
      to [short] (8, 2.5)
      to [short] (7, 2.5);
      \draw[Red, ultra thick] (8, 1.5) node[cross=8,color=Red, ultra thick,rotate=0] {};
\node at  (8.7, 1.5) {\large{$\hm{E_{J-}}$}};
\node at  (6.9, 1.5) {\large{$\hm{C_{J-}}$}};
\node at (3, 3.4) {\large{$\hm{\phi_{+}}$}};
     \draw (7,3) [Red, ultra thick]
     to [C, color=Red] (10,3);
     \draw (10,0) [Red, ultra thick]
     to [V, l, color=Red] (10,3);
     \draw (10,0) [Red, ultra thick]
     to [short] (7,0);
          \filldraw (7,3) circle (2pt);
          \filldraw (3,3) circle (2pt);
     \node at (7, 3.4) {\large{$\hm{\phi_{-}}$}};
                 \end{circuitikz}
    \caption{Basic circuit of the TCQ. The Josephson junctions can be substituted by flux tunable Squid loops allowing for the control of the Josephson energies $E_{J \pm}$.} 
    \label{tcq}
  \end{figure}
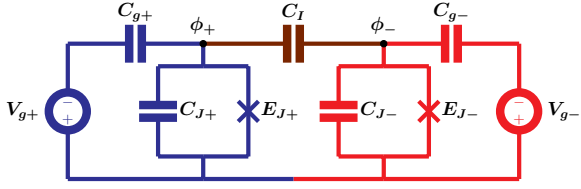

The TCQ, originally proposed by Gambetta et al. in \cite{tcq2011}, is essentially a system made up of two transmons that are coupled strongly by a capacitance, as shown in Fig. \ref{tcq}. The basic idea is that by encoding the qubit in the first two energy levels of this coupled system we can obtain a more flexible qubit, compared to a simple transmon, while still retaining a similar charge noise insensitivity. In particular, in the original paper it was shown how the TCQ \cite{tcq2011}, when coupled to a single resonator, can be operated in a configuration in which it is protected from the Purcell effect, but nonetheless still measurable via the standard dispersive readout technique. The independent tunability of the frequency of the TCQ and the coupling constant with the resonator via flux bias was more specifically analyzed in \cite{srinivasan2011}, while coherent control was shown in \cite{hoffman2011}. Recently, also the suppression of photon shot noise dephasing in a TCQ was demonstrated \cite{zhang2017}. In addition, a similar system employing inductive coupling between the transmons was proposed in \cite{diniz2013, dumur2015}.  \par
\begin{figure}
\vspace{0.5cm}
\centering
\begin{subfigure}[t]{0.2 \textwidth}
\includegraphics[scale=0.3]{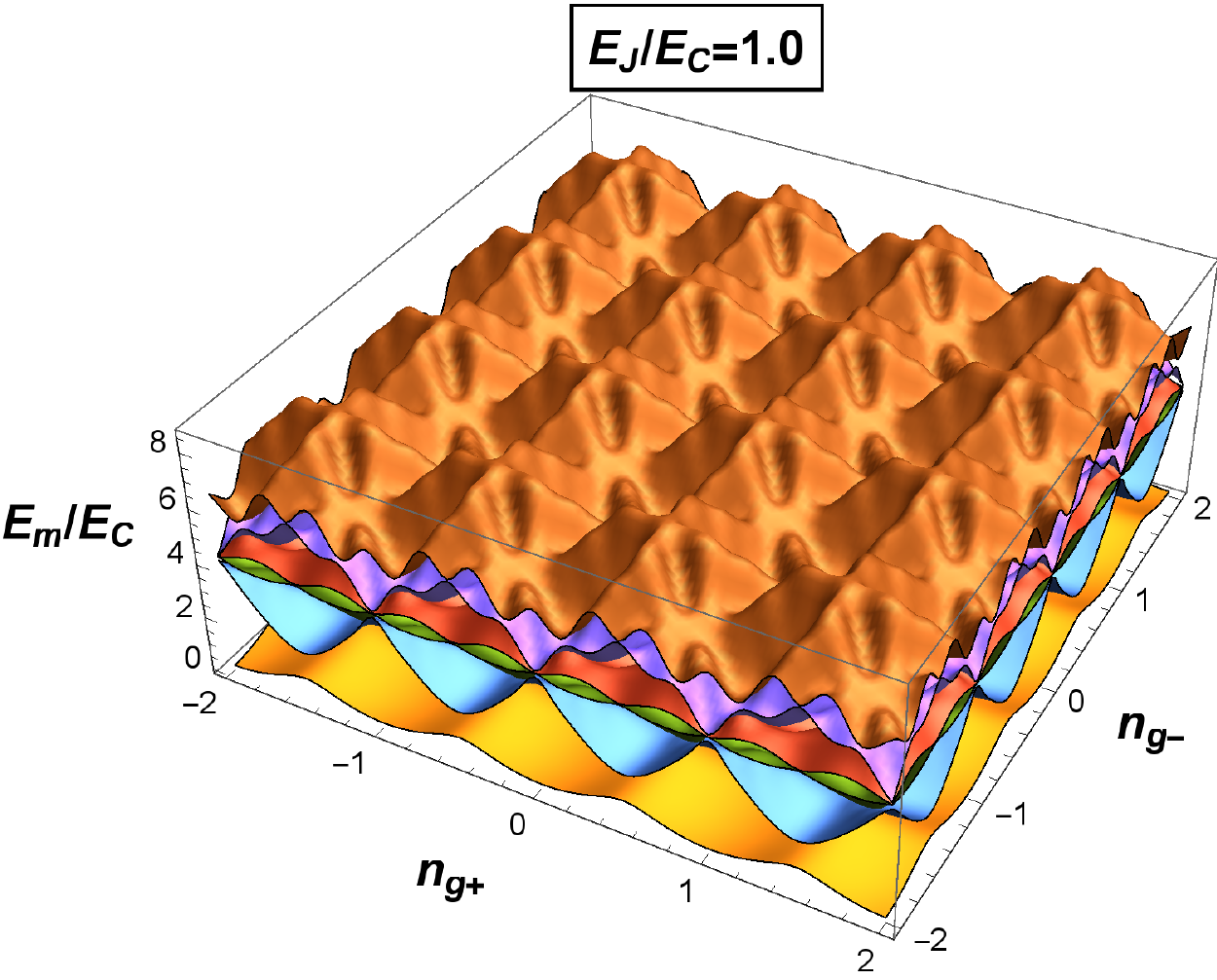}
\subcaption{}
\end{subfigure}
\begin{subfigure}[t]{0.2 \textwidth}
\includegraphics[scale=0.3]{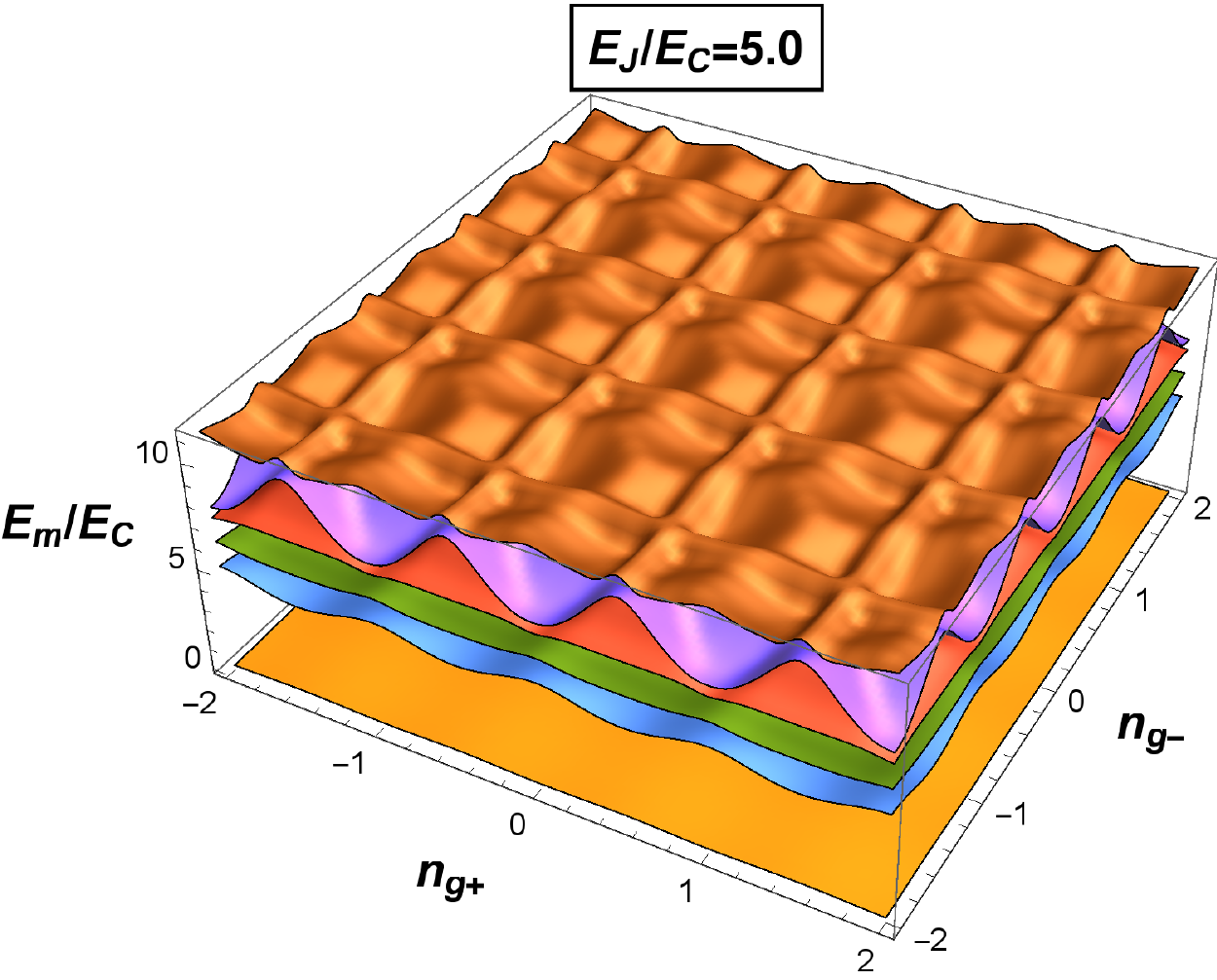}
\subcaption{}
\end{subfigure}
\begin{subfigure}[t]{0.2 \textwidth}
\includegraphics[scale=0.3]{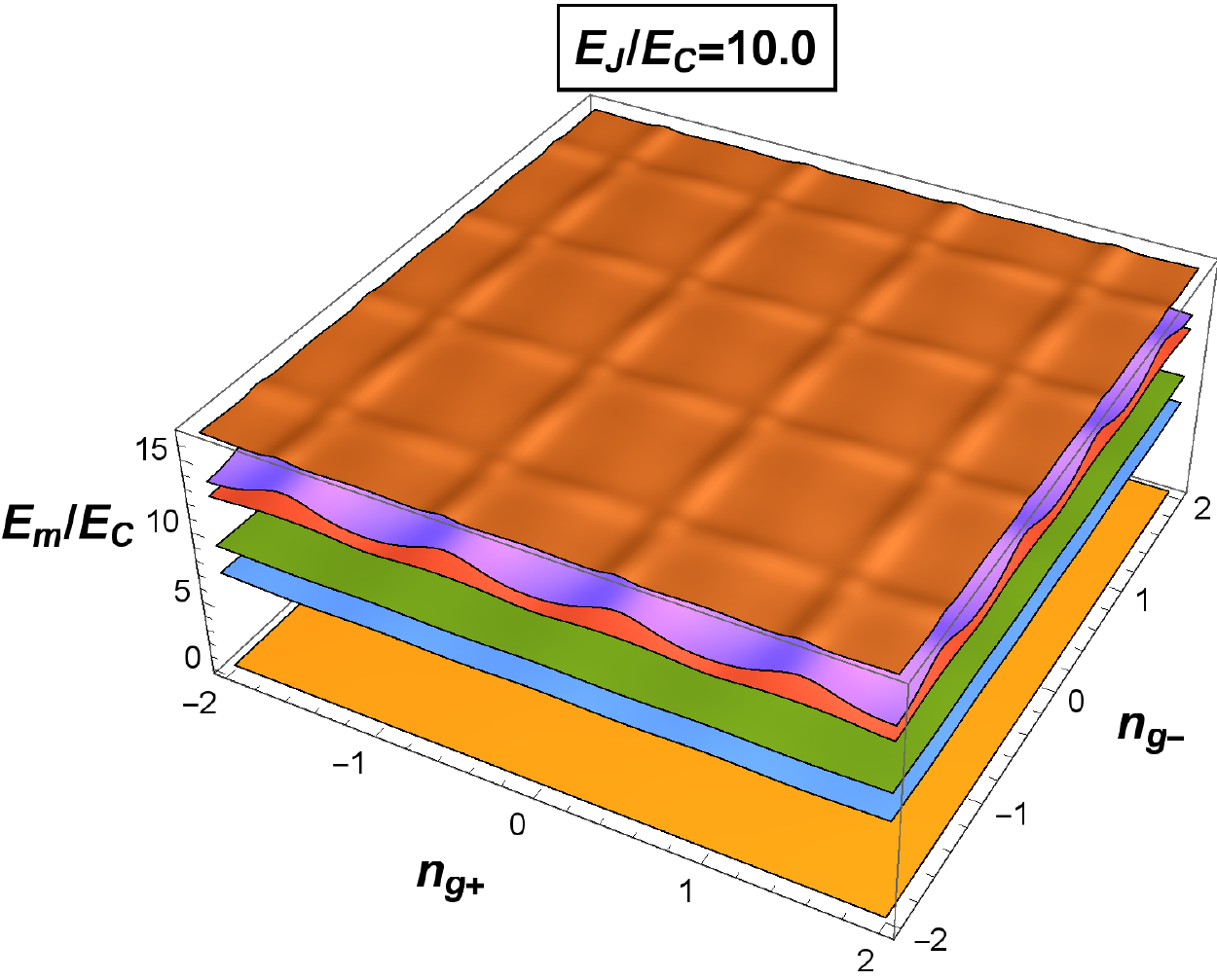}
\subcaption{}
\end{subfigure}
\begin{subfigure}[t]{0.2 \textwidth}
\includegraphics[scale=0.3]{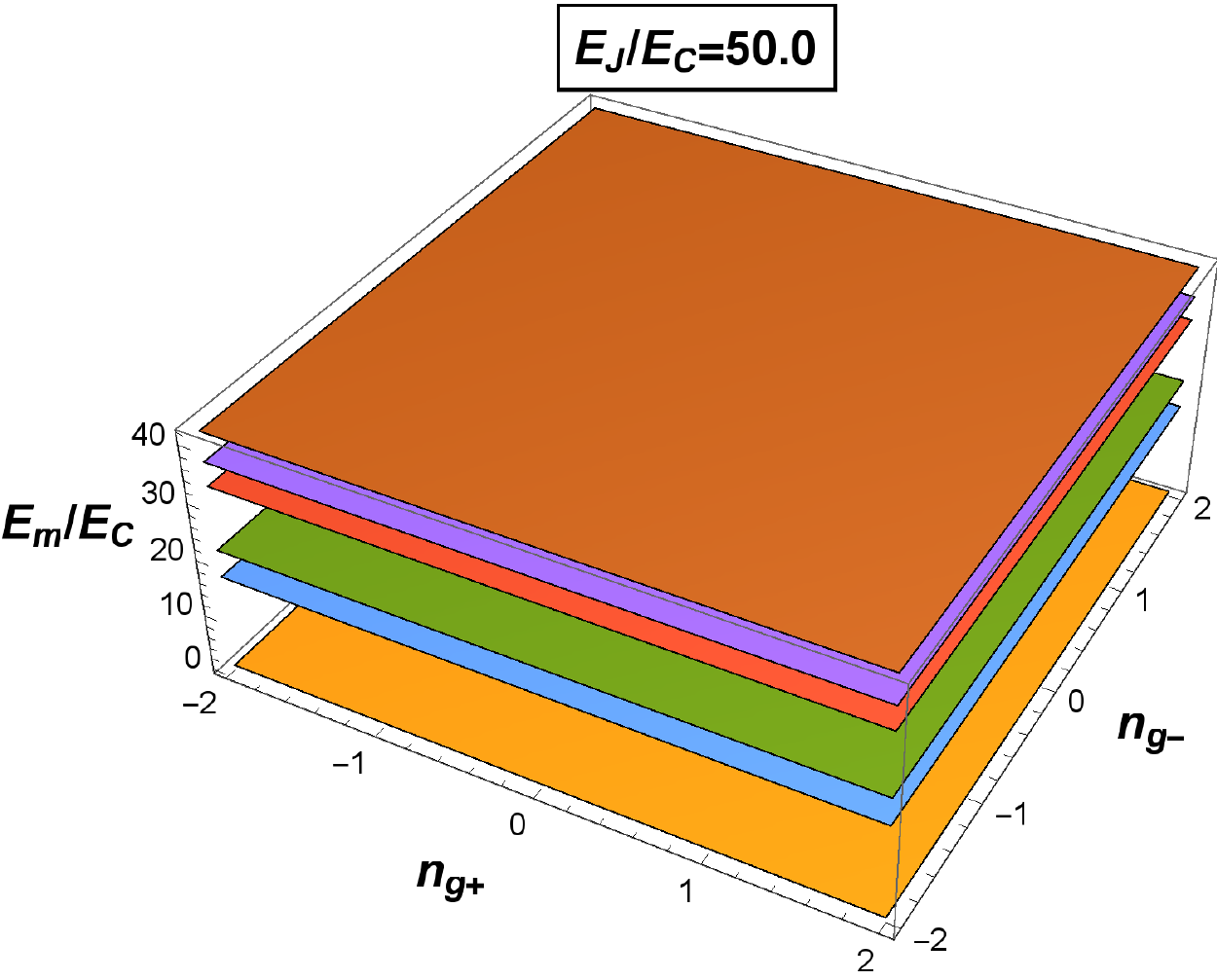}
\subcaption{}
\end{subfigure}
\caption{First 6 energy levels of the TCQ Hamiltonian Eq. \ref{Htcq1} as a function of the offset charges $n_{g+}$ and $n_{g-}$ ($m=\{0,1,2,3,4,5\}$). The zero of the energy is taken to be in each plot the minimum of the energies of the lowest level. The transmons are considered symmetric in this case and $E_I/E_C=-0.5$. The plots are obtained by direct numerical diagonalization of $H_{TCQ}$ by writing it in the charge basis and truncating the Hilbert space.}
\label{energyngTCQ}
\end{figure}

The Hamiltonian of the circuit in Fig. \ref{tcq} can be obtained via standard circuit quantization techniques \cite{devoret1997, bkd2004} as 
\begin{multline}
\label{Htcq1}
H_{TCQ}= \sum_{\pm} 4 E_{C \pm} (n_{\pm}-n_{g \pm})^2- \\ \sum_{\pm} E_{J \pm} \cos(\varphi_{\pm}) + 4 E_I (n_{+}-n_{g+}) (n_{-}-n_{g-}),
\end{multline}
where in the case in which all the capacitances are symmetric, \i.e., do not depend on the subscript $\pm$,  $E_{C +}=E_{C-}=E_{C}= e^2 (C_I+C_{\Sigma})/[2 (C_{\Sigma}^2+2 C_I C_{\Sigma})]$ and $E_I=-2 E_C C_I/(C_I+C_{\Sigma})$, with $C_{\Sigma}=C+C_g$. Thus, $E_I$ can be varied from $0$ to $-2 E_C$ by varying the interaction capacitance $C_I$. In Eq. \ref{Htcq1} we also introduced the reduced gate charges $n_{g \pm}= -C_{g \pm} V_{g \pm}/(2 e)$. If the ratios $E_{J \pm}/E_{C \pm}$ are sufficiently large we expect the energy levels of the TCQ to be basically independent of the parameters $n_{g \pm}$, \i.e., to show charge noise insensitivity, as with the transmon. This is shown explicitly in Fig. \ref{energyngTCQ}. When we operate the TCQ in this regime we can effectively neglect the reduced gate charges in the Hamiltonian Eq. \ref{Htcq1}. In addition, we can also expand each Josephson potential up to fourth order in $\varphi_{\pm}$ \citep{tcq2011} (as in Eq. \ref{Htransmon}). After introducing annihilation and creation operators for each transmon mode $b_{\pm}$ and $b_{\pm}^{\dagger}$ and neglecting fast rotating terms, we obtain the effective Hamiltonian of the TCQ as two coupled Duffing oscillators
\begin{multline}
\label{HTCQeff1}
H_{TCQeff}= \sum_{\pm}  \omega_{\pm} b_{\pm}^{\dagger} b_{\pm}+ \frac{\delta_{\pm}}{2} b_{\pm}^{\dagger}b_{\pm}^{\dagger}b_{\pm} b_{\pm} + \\  J (b_{+} b_{-}^{\dagger}+b_{+}^{\dagger} b_{-}),
\end{multline}  
with $\omega_{\pm}=\sqrt{8 E_{j \pm} E_{C \pm}}-E_{C \pm}$, $\delta_{\pm}=-E_{C \pm}$, $J=1/(\sqrt{2}) E_I (E_{J+}/E_{C+})^{1/4}(E_{J-}/E_{C-})^{1/4}$, and commutation relations $[b_{\pm}, b_{\pm}^{\dagger}]=1$. 
The Hamiltonian Eq. \ref{HTCQeff1} can be approximately diagonalized via the unitary transformation 
\begin{equation}
\label{Utcq}
U_{TCQ}= \exp[\lambda(b_+ b_-^{\dagger}-b_+^{\dagger} b_-)],
\end{equation}
with 
\begin{equation}
\label{lambda}
\lambda= 1/2 \arctan(-2 J/\zeta),
\end{equation}
where we defined $\zeta= \omega_+-\omega_- -2(\delta_+-\delta-)$. Applying this transformation to each of the annihilation operators we get \footnote{The tildes are used to point out that they are annihilation operators in the new basis.}
\begin{subequations}
\begin{equation}
U_{TCQ} b_{+} U_{TCQ}^{\dagger} = \cos (\lambda) \tilde{b}_{+} + \sin (\lambda) \tilde{b}_{-},
\end{equation}
\begin{equation}
U_{TCQ} b_{-} U_{TCQ}^{\dagger} = -\sin (\lambda) \tilde{b}_{+} + \cos (\lambda) \tilde{b}_{-},
\end{equation}
\end{subequations} 
from which we obtain the Hamiltonian 
\begin{multline}
\label{HTCQeff}
\tilde{H}_{TCQeff}= U_{TCQ} H_{TCQ} U_{TCQ}^{\dagger} \approx \\ \sum_{\pm}  \tilde{\omega}_{\pm} \tilde{b}_{\pm}^{\dagger} \tilde{b}_{\pm}+\frac{\tilde{\delta}_{\pm}}{2} \tilde{b}_{\pm}^{\dagger} \tilde{b}_{\pm}^{\dagger} \tilde{b}_{\pm} \tilde{b}_{\pm}^{\dagger}+ \\ \tilde{\delta}_{c} \tilde{b}_{+}^{\dagger} \tilde{b}_{+} \tilde{b}_{-}^{\dagger} \tilde{b}_{-},
\end{multline}
where $\tilde{\omega}_{\pm}= (\omega_+ + \omega_-)/2 \pm (\omega_+- \omega_-) \cos (2 \lambda)/2 \mp J \sin(2 \lambda)$, $\tilde{\delta}_{\pm}= (\delta_+ + \delta_-)(1+\cos^2(2 \lambda))/2 \pm (\delta_+-\delta_-)\cos(2 \lambda)/2$ and $\tilde{\delta}_c= (\delta_+ + \delta_-) \sin^2 (2 \lambda)/2$. In this procedure, we have assumed $\lvert \delta_{\pm}/(\tilde{\omega}_+-\tilde{\omega}_-) \rvert \ll 1$. 
\subsection{TCQ coupled to two resonators}
\label{tcq2res}
\begin{figure}
\centering
\begin{tikzpicture}[scale=0.8, every node/.style={transform shape}]
\draw [ultra thick, BrickRed] (-1,0)--(1,0);
\node at (0, -0.3) {$\large{\hm{\ket{0_{+ } 0_{-}}}}$};
\draw [ultra thick, Gray] (-1,4.5)--(1,4.5);
\node at (0, 5.3) {$\large{\hm{\ket{1_{+} 1_{-}}}}$};
\node at (0, 4.8) {$\large{\hm{\tilde{\omega}_{+}+\tilde{\omega}_{-}+\tilde{\delta}_{c}}}$};
\draw[ultra thick, BrickRed] (-2.5, 2)--(-0.5, 2);
\node at (-3.2, 2) {$\large{\hm{\ket{0_{+} 1_{-}}}}$};
\node at (-0.3, 2.3) {$\large{\hm{\tilde{\omega}_{-}}}$};
\draw[ultra thick, Gray] (0.5, 3)--(2.5,3);
\node at (3.2, 3) {$\large{\hm{\ket{1_{+} 0_{-}}}}$};
\node at (0.3, 3.3) {$\large{\hm{\tilde{\omega}_{+}}}$};
\draw [ultra thick, Gray] (-4,3.5)--(-2,3.5);
\node at (-4.7,3.5) {$\large{\hm{\ket{0_{+ } 2_{-}}}}$};
\node at (-3,3.8) {$\large{\hm{2\tilde{\omega}_{-}+\tilde{\delta}_{-}}}$};
\draw[ultra thick, Gray] (2, 5.5)--(4,5.5);
\node at (4.7, 5.5) {$\large{\hm{\ket{2_{+} 0_{-}}}}$};
\node at (3, 5.8) {$\large{\hm{2\tilde{\omega}_{+}+\tilde{\delta}_{+}}}$};
\draw [thick, triangle 45-triangle 45, DarkGreen, dashed] (-0.5, 0)--(-1.5, 2);
\draw [thick, triangle 45-triangle 45, DarkGreen] (+0.5, 0)--(1.5, 3);
\draw [thick, triangle 45-triangle 45, Bittersweet, dashed] (-0.5+0.4, 0)--(-1.5+0.4, 2);
\draw [thick, triangle 45-triangle 45, Bittersweet] (+0.5-0.4, 0)--(1.5-0.4, 3);
\draw [thick, triangle 45-triangle 45, DarkGreen] (-1.5, 2)--(-0.7, 4.5);
\draw [thick, triangle 45-triangle 45, Bittersweet] (-1.5+0.4, 2)--(-0.7+0.4, 4.5);
\draw [thick, triangle 45-triangle 45, DarkGreen, dashed] (1.5, 3)--(+0.7, 4.5);
\draw [thick, triangle 45-triangle 45, Bittersweet, dashed] (1.5-0.4, 3)--(+0.7-0.4, 4.5);
\draw [thick, triangle 45-triangle 45, DarkGreen, dashed] (-2.2, 2)--(-3, 3.5);
\draw [thick, triangle 45-triangle 45, Bittersweet, dashed] (-2.2+0.4, 2)--(-3+0.4, 3.5);
\draw [thick, triangle 45-triangle 45, DarkGreen] (+2.2, 3)--(3, 5.5);
\draw [thick, triangle 45-triangle 45, Bittersweet] (+2.2-0.4, 3)--(3-0.4, 5.5);
\end{tikzpicture}
\caption{Energy levels and transition diagrams for the TCQ. The first two levels in red are those in which we encode the qubit. The green and orange arrows denote respectively transitions that might be caused by resonator 1 and 2. The solid arrows denote ``$+$" transitions, while the dashed arrows ``$-$" transitions.}
\label{tcqLevelsTrans}
\end{figure}
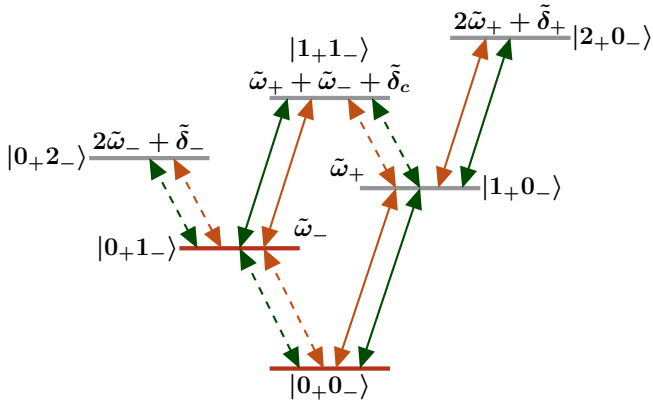

To return to our parity-measurement setup, we now analyze the case in which each transmon mode composing the TCQ is coupled linearly to two bosonic modes. This means that by modelling the TCQ directly as two coupled Duffing oscillators the total Hamiltonian would read
\begin{multline}
H= H_{TCQeff}+ \sum_{i=1}^2 \omega_i a_i^{\dagger} a_i + \sum_{i=1}^2 \sum_{\pm} g_{i \pm}\bigl(a_i b_{\pm}^{\dagger}+\mathrm{H.c.} \bigr),
\end{multline}
where for now we treat the bare coupling coefficients $g_{i \pm}$ as free real parameters. Applying the unitary transformation in Eq. \ref{Utcq} to this Hamiltonian we finally get
\begin{multline}
\label{Htcq2res}
\tilde{H}= U_{TCQ} H U_{TCQ}^{\dagger} \approx  \\ \tilde{H}_{TCQeff} + \sum_{i=1}^2 \omega_i a_i^{\dagger} a_i + \sum_{i=1}^2 \sum_{\pm} \tilde{g}_{i \pm}\bigl(a_i \tilde{b}_{\pm}^{\dagger} + \mathrm{H.c.} \bigr),
\end{multline}
where now in this new basis we have effective coupling parameters that are a linear combination of the bare ones. In particular
\begin{equation} 
\label{geff}
\tilde{g}_{i \pm}= g_{i +} \cos(\lambda) \mp g_{i -} \sin(\lambda).
\end{equation} This is one of the central features of the TCQ, and the one that we will exploit. We see in Fig. \ref{tcqLevelsTrans} the first six levels of the TCQ and the possible transitions that might be induced the linear interaction with the resonators. Clearly, in the general case, the two resonators can cause exactly the same transitions, but we will be interested in a more particular configuration.\par
At this point we want to procede like we did for the transmon, obtaining an effective Hamiltonian for the first two levels of the TCQ, namely $\ket{0_+ 0_-}$ and $\ket{1_+ 0_-}$, in the dispersive regime of the qubit-resonator interactions. We start by projecting the Hamiltonian Eq. \ref{Htcq2res} onto the subspace spanned by the first six levels of the TCQ. These are the relevant states that one must consider in order to get the correct dispersive shifts for the first excited manifold, namely $\ket{0_+ 1_-}$ and $\ket{1_+ 0_-}$.  We get
\begin{widetext}
\begin{multline}
\label{Htcq2resP}
\tilde{H}_{p}= \tilde{\omega}_+ \ket{1_+ 0_-} \bra{1_+ 0_-} + \tilde{\omega}_- \ket{0_+ 1_-} \bra{0_+ 1_-} + (2 \tilde{\omega}_+ + \tilde{\delta}_+) \ket{2_+ 0_-} \bra{2_+ 0_-} + (2 \tilde{\omega}_- + \tilde{\delta}_-) \ket{0_+ 2_-} \bra{0_+ 2_-} + \\ (\tilde{\omega}_+ + \tilde{\omega}_- +  \tilde{\delta}_c) \ket{1_+ 1_-}\bra{1_+ 1_-} +\sum_{i=1}^2 \omega_i a_i^{\dagger} a_i + \\
\sum_{i=1}^2 \biggl \{\tilde{g}_{i +} \bigl[a_i \bigl(\ket{1_+ 0_-}\bra{0_+ 0_-}+\ket{1_+ 1_-} \bra{0_+ 1_-} + \sqrt{2} \ket{2_+ 0_-} \bra{1_+ 0_-} + \mathrm{H.c.} \bigr) \bigr]+\\ \tilde{g}_{i -} \bigl[a_i \bigl(\ket{0_+ 1_-}\bra{0_+ 0_-}+\ket{1_+ 1_-} \bra{1_+ 0_-} + \sqrt{2} \ket{0_+ 2_-} \bra{0_+ 1_-} + \mathrm{H.c.} \bigr) \bigr] \biggr \}.
\end{multline} 
\end{widetext}
The procedure is now similar to what has been carried out in Sec. \ref{sec3qstransmon}. First of all let us define the detunings $\tilde{\Delta}_{\pm i }= \tilde{\omega}_{\pm}-\omega_i$. Then, we consider all the interactions to be in the dispersive regime, which amounts to assuming $\lvert  \tilde{g}_{i \pm}/\tilde{\Delta}_{\pm i} \rvert \ll 1$, $\lvert \sqrt{2} \tilde{g}_{i \pm}/(\tilde{\Delta}_{\pm}+\tilde{\delta}_{\pm}) \rvert \ll 1$ and $\lvert \tilde{g}_{i \pm}/(\tilde{\Delta}_{\pm}+\tilde{\delta}_c)\rvert \ll 1$. The generator $S$ of the Schrieffer-Wolff transformation $D=\exp[S-S^{\dagger}]$ that removes the interaction between the TCQ  and the resonators is found to be
\begin{multline}
S= \sum_{i=1}^2 a_i \biggl \{\frac{\tilde{g}_{i +}}{\tilde{\Delta}_{+ i}} \ket{1_+ 0_-}\bra{0_+ 0_-} + \\ \frac{\tilde{g}_{i +}}{\tilde{\Delta}_{+ i}+\tilde{\delta}_c} \ket{1_+ 1_-} \bra{0_+ 1_-} +  \frac{\sqrt{2} \tilde{g}_{i +}}{\tilde{\Delta}_{+i}+\tilde{\delta}_+} \ket{2_+ 0_-} \bra{1_+ 0_-} + \\
\frac{\tilde{g}_{i -}}{\tilde{\Delta}_{- i}} \ket{0_+ 1_-}\bra{0_+ 0_-} +  \frac{\tilde{g}_{i -}}{\tilde{\Delta}_{- i}+\tilde{\delta}_c} \ket{1_+ 1_-} \bra{1_+ 0_-} + \\ \frac{\sqrt{2} \tilde{g}_{i -}}{\tilde{\Delta}_{-i}+\tilde{\delta}_-} \ket{0_+ 2_-} \bra{0_+ 1_-} \biggl \}.
\end{multline}
The result of this transformation is given in Eq. \ref{tcqDispTot} in Appendix \ref{appDispTCQ}. Projecting Eq. \ref{tcqDispTot} onto the two-dimensional subspace spanned by $\{\ket{0_+ 0_-}, \ket{0_+ 1_-}$, which thus encodes our qubit, we obtain
\begin{multline}
\label{Htcq2res1}
\tilde{H} = \tilde{\Omega}_- \ket{0_+ 1_-}\bra{0_+ 1_-} + \sum_{i=1}^2 \bigl [\omega_i + \\ \chi_{i, 0_+ 1_-}\ket{0_+1_-}\bra{0_+ 1_-}- \chi_{i, 0_+ 0_-} \ket{0_+ 0_-} \bra{0_- 0_+} \bigr] a_i^{\dagger}a_i  + \\
\bigr [\chi_{12, 0_+ 1_-}\ket{0_+ 1_-}\bra{0_+ 1_-} - \\ \chi_{12, 0_+ 0_-} \ket{0_+ 0_-} \bra{0_+ 0_-} \bigl] \bigl(a_1 a_2^{\dagger}\,+ \,\mathrm{H.c.} \bigr),
\end{multline}
where we denoted by $\chi_{i, 0_+ 1_-}$ and $\chi_{i, 0_+ 0_-}$ the dispersive shifts of the frequency of resonator $i$ caused by state $\ket{0_+ 1_-}$ and $\ket{0_+ 0_-}$ respectively. In addition, $\chi_{12, 0_+ 1_-}$ represents the coupling coefficient between the two resonators when the qubit is in the state $\ket{0_+ 1_-}$, while $\chi_{12,0_+ 0_-}$ if the state is $\ket{0_+ 0_-}$. These parameters are given by the following formulas
\begin{subequations}
\begin{equation}
\chi_{i,0_+ 1_-} = \frac{\tilde{g}_{i -}^2}{\tilde{\Delta}_{i -}} - \frac{(\sqrt{2} \tilde{g}_{i-})^2}{\tilde{\Delta}_{i -} +\tilde{\delta}_-} - \frac{\tilde{g}_{i +}^2}{\tilde{\Delta}_{i+}+\tilde{\delta}_c}, 
\end{equation}
\begin{equation}
\chi_{i, 0_+ 0_-}= \frac{\tilde{g}_{i +}^2}{\tilde{\Delta}_{i+}} +\frac{\tilde{g}_{i -}^2}{\tilde{\Delta}_{i-}}, 
\end{equation}
\begin{multline}
\chi_{12, 0_+ 1_-}= \frac{\tilde{g}_{1 -} \tilde{g}_{2 -}}{2} \biggl (\frac{1}{\tilde{\Delta}_{1 -}} +\frac{1}{\tilde{\Delta}_{2 -}} \biggr) - \\ \frac{(\sqrt{2}\tilde{g}_{1 -})(\sqrt{2} \tilde{g}_{2 -})}{2} \biggl (\frac{1}{\tilde{\Delta}_{1 -}+\tilde{\delta}_-} +\frac{1}{\tilde{\Delta}_{2 -}+\tilde{\delta}_-} \biggr)- \\ \frac{\tilde{g}_{1 +} \tilde{g}_{2 +}}{2} \biggl (\frac{1}{\tilde{\Delta}_{1 +}+\tilde{\delta}_c} +\frac{1}{\tilde{\Delta}_{2 +}+\tilde{\delta}_c} \biggr)   
\end{multline}
\begin{multline}
\chi_{12,0_+ 0_-} = \frac{\tilde{g}_{1 +} \tilde{g}_{2 +}}{2} \biggl (\frac{1}{\tilde{\Delta}_{1 +}} +\frac{1}{\tilde{\Delta}_{2 +}} \biggr) + \\ \frac{\tilde{g}_{1 -} \tilde{g}_{2 -}}{2} \biggl (\frac{1}{\tilde{\Delta}_{1 -}} +\frac{1}{\tilde{\Delta}_{2 -}} \biggr),
\end{multline}
\end{subequations}
$i={1,2}$. Finally, expressing the projectors in terms of the Pauli $\sigma^z$ operator, namely $\ket{0_+ 1_-}\bra{0_+ 1_-}= (1+\sigma^z)/2$ and $\ket{0_+ 0_-}\bra{0_+ 0_-}= (1-\sigma^z)/2$, we rewrite the Hamiltonian Eq. \ref{Htcq2res1} as
\begin{multline}
\tilde{H} = \frac{\tilde{\Omega}_-}{2} \sigma^z + \sum_{i=1}^2
 \bigl (\overline{\omega}_i + \chi_{i} \sigma^z \bigr) a_{i}^{\dagger}a_i + \\
\biggl [\bar{\chi}_{12} + \chi_{12} \sigma^z \biggr] \bigl(a_1 a_2^{\dagger} +\mathrm{H.c.}, \bigr)  
 \end{multline}
which is essentially the same as Eq. \ref{HteffTLS} obtained for the transmon, but with modified parameters
 \begin{subequations}
 \label{chisTCQ}
 \begin{equation}
 \bar{\omega}_i = \omega_i + \frac{\chi_{i, 0_+ 1_-}-\chi_{i,0_+ 0_-}}{2},
 \end{equation}
 \begin{equation}
 \chi_{i} = \frac{\chi_{i, 0_+ 1_-}+\chi_{i,0_+ 0_-}}{2},
 \end{equation}
 \begin{equation}
 \bar{\chi}_{12} = \frac{\chi_{12, 0_+ 1_-}-\chi_{12,0_+ 0_-}}{2},
 \end{equation}
\begin{equation}
 \chi_{12} = \frac{\chi_{12, 0_+ 1_-}+\chi_{12,0_+ 0_-}}{2},
\end{equation}
 \end{subequations}
 $i={1,2}$. \par
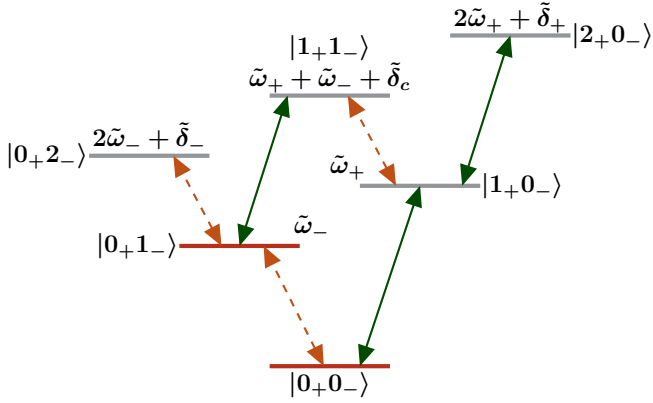
\begin{figure}
\flushleft
\begin{tikzpicture}[scale=0.8, every node/.style={transform shape}]
\draw [ultra thick, BrickRed] (-1,0)--(1,0);
\node at (0, -0.3) {$\large{\hm{\ket{0_{+ } 0_{-}}}}$};
\draw [ultra thick, Gray] (-1,4.5)--(1,4.5);
\node at (0, 5.3) {$\large{\hm{\ket{1_{+} 1_{-}}}}$};
\node at (0, 4.8) {$\large{\hm{\tilde{\omega}_{+}+\tilde{\omega}_{-}+\tilde{\delta}_{c}}}$};
\draw[ultra thick, BrickRed] (-2.5, 2)--(-0.5, 2);
\node at (-3.2, 2) {$\large{\hm{\ket{0_{+} 1_{-}}}}$};
\node at (-0.3, 2.3) {$\large{\hm{\tilde{\omega}_{-}}}$};
\draw[ultra thick, Gray] (0.5, 3)--(2.5,3);
\node at (3.2, 3) {$\large{\hm{\ket{1_{+} 0_{-}}}}$};
\node at (0.3, 3.3) {$\large{\hm{\tilde{\omega}_{+}}}$};
\draw [ultra thick, Gray] (-4,3.5)--(-2,3.5);
\node at (-4.7,3.5) {$\large{\hm{\ket{0_{+ } 2_{-}}}}$};
\node at (-3,3.8) {$\large{\hm{2\tilde{\omega}_{-}+\tilde{\delta}_{-}}}$};
\draw[ultra thick, Gray] (2, 5.5)--(4,5.5);
\node at (4.7, 5.5) {$\large{\hm{\ket{2_{+} 0_{-}}}}$};
\node at (3, 5.8) {$\large{\hm{2\tilde{\omega}_{+}+\tilde{\delta}_{+}}}$};
\draw [thick, triangle 45-triangle 45, DarkGreen] (+0.5, 0)--(1.5, 3);
\draw [thick, triangle 45-triangle 45, Bittersweet, dashed] (-0.5+0.4, 0)--(-1.5+0.4, 2);
\draw [thick,  triangle 45-triangle 45, DarkGreen] (-1.5, 2)--(-0.7, 4.5);
\draw [thick,  triangle 45-triangle 45, Bittersweet, dashed] (1.5-0.4, 3)--(+0.7-0.4, 4.5);
\draw [thick,  triangle 45-triangle 45, Bittersweet, dashed] (-2.2+0.4, 2)--(-3+0.4, 3.5);
\draw [thick,  triangle 45-triangle 45, DarkGreen] (+2.2, 3)--(3, 5.5);
\end{tikzpicture}
\caption{Energy levels and transitions diagram for the case $\tilde{g}_{1 -} = 0$ and $\tilde{g}_{2 +}=0$. The green and orange arrows denote respectively transitions that might be caused by resonator 1 and 2. Since the two resonators cannot excite the same transitions they do no interact with each other in the dispersive regime of the interactions.}
\label{tcqLevelsChi120}
\end{figure}

 
 From the previous formulas it is clear how the TCQ offers more freedom in the choice of the parameters compared to the standard transmon. In particular, it is possible to cancel the interaction between the resonators mediated by the qubit, while still retaining a qubit-state dependent frequency shift on both resonators.  This condition is achieved in the situation in which $\tilde{g}_{1 -} = 0$ and $\tilde{g}_{2 +}=0$ (or symmetrically  $\tilde{g}_{1 +} = 0$ and $\tilde{g}_{2 -}=0$) for which both $\bar{\chi}_{12}$ and $\chi_{12}$ given in Eqs. \ref{chisTCQ} are clearly zero. Nevertheless, $\chi_1$ and $\chi_2$ are not zero and given by
 \begin{subequations}
 \label{chiTCQ}
 \begin{equation}
 \chi_1 =  \frac{\tilde{g}_{1+}^2}{2} \frac{\tilde{\delta}_c}{\tilde{\Delta}_{1+}(\tilde{\Delta}_{1+}+ \tilde{\delta}_c)},
 \end{equation}
 \begin{equation}
 \label{chi2TCQ}
 \chi_2= \tilde{g}_{2-}^2 \frac{\tilde{\delta}_-}{\tilde{\Delta}_{2-}(\tilde{\Delta}_{2-} + \tilde{\delta}_-)}.
 \end{equation}
 \end{subequations} 
 The energy levels and transition diagram for this configuration is shown in Fig. \ref{tcqLevelsChi120}, from which we also understand the reason why the resonators do not interact with each other via the TCQ. In fact, in this case resonator $1$ is able to generate only ``$+$" transitions, while resonator $2$ only ``$-$" transitions. Consequently, resonator 1 and 2 are not capable to cause simultaneously the same transition and therefore the exchange of a photon between the resonators via a ``virtual" transition of the TCQ cannot happen. This reasoning is actually general and not limited to the TCQ. Thus, any system in which two bosonic modes cannot cause the same transitions would not produce an interaction between the resonators (at least in the dispersive regime). Another candidate system for satisfying this condition might be the trimon \cite{trimon}, which in turn can be viewed as an evolution of the TCQ. In addition, as we see from Eqs. \ref{chiTCQ} the anharmonicities $\tilde{\delta}_-$ and $\tilde{\delta}_c$ guarantee a non zero dispersive shift for both resonators. This would not be true for a system of two transversally coupled two-level systems in which also these dispersive shifts would be zero \footnote{Notice that one cannot obtain this case just by taking $\tilde{\delta} \rightarrow +\infty$ and $\tilde{\delta} \rightarrow \infty$ for the TCQ, as the approximate diagonalization of the TCQ Hamiltonian we carried out requires small anharmonicities.}. Finally, we point out that in the configuration shown in Fig. \ref{tcqLevelsChi120} the TCQ would not be Purcell protected, since it can decay via the interaction with resonator $2$.      \par 
A possible implementation of this condition might be to set $\lambda= \pi/4$, which means that the coupling coefficient between the bare transmons $J$ is much larger than their detuning. Additionally, we require the bare $g_{i \pm}$ coupling coefficients to have a sign flip, that is to say we set
\begin{subequations}
\label{signFlip}
\begin{equation}
g_{1+}= -g_{1-}=g_1, 
\end{equation}
\begin{equation}
g_{1 +} = +g_{2 -}=g_2,
\end{equation}
\end{subequations}
from which we readily obtain the effective coefficients in the diagonalized basis $\tilde{g}_{1-}= \tilde{g}_{2 +}=0$ and $\tilde{g}_{1+}= \sqrt{2}g_1$, $\tilde{g}_{2+}= \sqrt{2} g_2$. \par
Implementing this sign flip might not be trivial and it is the topic of the next subsection.
\subsection{Achieving zero $\chi_{12}$}
\label{achZerochi12}
We here tackle the problem of achieving the sign flip of the bare coupling coefficients described in the previous subsection, in order to cancel the quantum switch term. In order to do this, we need to exploit the distributed character of a resonator. In particular, we will consider the usual case in which the resonator is implemented as a coplanar waveguide resonator, which is the original proposal for the circuit QED architecture \cite{blais2004}. This allows an analytical treatment and conveys the main idea for cancelling the $\chi_{12}$ term. However, similar reasonings apply also to other kinds of non-lumped resonators, such as 3D microwave cavities. \par 
In Appendix \ref{appTransmonTL} we obtain the generic coupling coefficient of a transmon capacitively coupled to a transmission line resonator of length $L$ at a certain position $x_j$. Considering only the second mode of the resonator, we obtain the coupling coefficient to be of the form
\begin{equation}
g(x_j)= g_{max} \cos \biggl(\frac{2 \pi}{L} x_j \biggr),
\end{equation}  
where the coupling parameter $g_{max}$ to a generic resonator mode is given in detail in Eq. \ref{gxj}. Hence, we see that depending on the position of the transmon we can access regions with positive and negative signs of the coupling parameter.
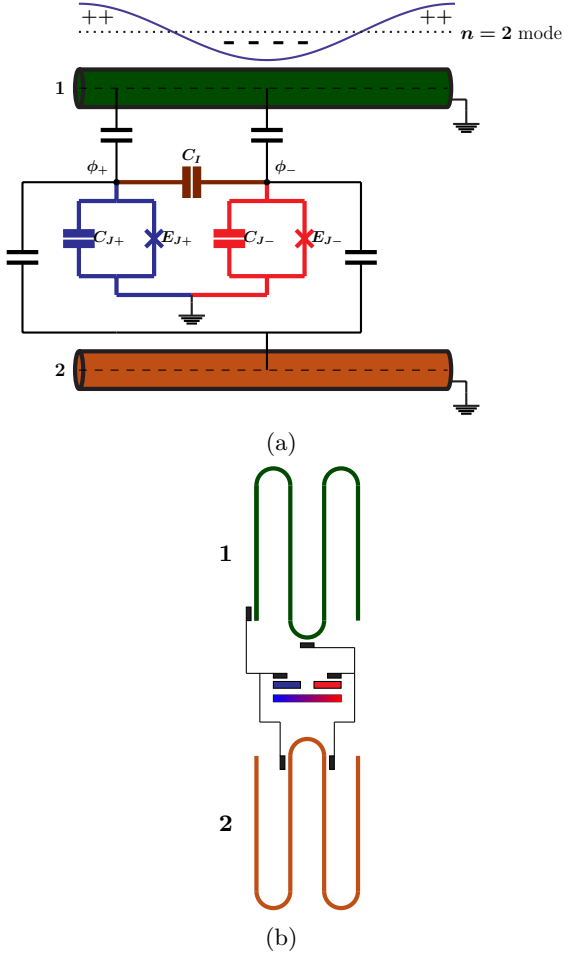
\begin{figure}
\vspace{5mm}
 \begin{subfigure}[b]{0.45\textwidth}
\centering
   \begin{circuitikz}[american voltages, scale=0.5, every node/.style={transform shape}]
     \draw[Blue, ultra thick] (3,3)
      to [short] (3,2.5)
      to [short] (2,2.5)
      to[C, color=Blue] (2,0.5) 
      to [short] (3,0.5)
      to [short] (3,0);
       \draw[Blue, ultra thick] (3, 2.5)
      to [short] (4,2.5)
      to [short] (4, 0.5)
      to [short] (3,0.5);
     \draw[Blue, ultra thick] (4, 1.5) node[cross=8,color=Blue, ultra thick,rotate=0] {};
\node at  (4.6, 1.5) {\large{$\hm{E_{J+}}$}};
\node at  (2.8, 1.5) {\large{$\hm{C_{J+}}$}};
\node at  (5, 3.7) {\large{$\hm{C_{I}}$}};
      \draw[Brown, ultra thick] (3,3)
      to [C, color=Brown] (7,3);
      \draw[Red, ultra thick] (7,3)
      to [short] (7,2.5)
      to [short] (6, 2.5)
      to [C, color=Red] (6, 0.5)
      to [short] (7,0.5)
      to [short] (7,0)
      to [short] (5,0);
      \draw[Blue, ultra thick] (5,0)
      to [short] (3,0);
      \draw[Black, thick] (11.8,5.2)
      to [short] (12.3, 5.2)
      to (12.3,4.9) node[ground]{};
       \draw[Black, thick] (11.8,5.2-7.5)
      to [short] (12.3, 5.2-7.5)
      to (12.3,4.9-7.5) node[ground]{};
      \draw[Black, thick] (5,0)
      to (5,-0.2) node[ground]{};
      \draw[Red, ultra thick] (7, 0.5)
      to [short] (8,0.5)
      to [short] (8, 2.5)
      to [short] (7, 2.5);
      \draw[Red, ultra thick] (8, 1.5) node[cross=8,color=Red, ultra thick,rotate=0] {};
\node at  (8.6, 1.5) {\large{$\hm{E_{J-}}$}};
\node at  (6.8, 1.5) {\large{$\hm{C_{J-}}$}};
\node at (2.5, 3.4) {\large{$\hm{\phi_{+}}$}};
\node at (7.5, 3.4) {\large{$\hm{\phi_{-}}$}};
    
          \filldraw (7,3) circle (2pt);
           \filldraw (3,3) circle (2pt);
     
     \node at (7,5.5)(a) [cylinder,Black,fill=DarkGreen, shape border rotate=180, draw,minimum height=10cm,minimum width=1cm, ultra thick]
{};
       \node at (7,-2)(a) [cylinder,Black,fill=Bittersweet, shape border rotate=180, draw,minimum height=10cm,minimum width=1cm, ultra thick]
{};
      \draw[thick] (3,3)
      to [C] (3,5.5);
      \draw[thick] (7,3)
      to [C] (7,5.5);
      \draw  [thick] (3,3) 
      to [short] (0.5,3)
      to [C] (0.5,-1)
      to [short] (3,-1);
      \draw [thick] (7,3)
      to [short] (9.5,3)
      to [C] (9.5,-1)
       to [short] (3,-1);
       \draw[thick](7,-1)
      to [short] (7,-2);
      \draw [dashed] (7-5,5.5) -- (7+4.8,5.5);
       \draw [dashed] (7-5,-2) -- (7+4.8,-2);
        \node at (1.5, 5.5) {\Large{$\hm{1}$}};
        \node at (1.5, -2) {\Large{$\hm{2}$}};
         \node at (2.5, 7.4) {\textbf{\Large{++}}};
           \node at (11.5, 7.4) {\textbf{\Large{++}}};
           \node at (7, 6.7) {\textbf{\Huge{- - - -}}};
           \node at (13.5, 7) {\Large{$\hm{n=2}$ mode}};
        \draw[scale=0.5,smooth,variable=\x,BlueViolet, samples at={4, 4.25,..., 24}, thick] plot ({\x},{14+1.5*cos(18*(\x-4))});
          \draw[scale=0.5,smooth,variable=\x,Black,dotted, samples at={4, 4.25,..., 24}, thick] plot ({\x},{14});
                       \end{circuitikz}
                       \caption{•}
                       \end{subfigure}
                        \begin{subfigure}[b]{0.4\textwidth}
                        \centering
  \begin{tikzpicture}[scale=0.9]
 \draw[ultra thick, DarkGreen] (-0.25,3)--(-0.25,1);
  \draw[ultra thick, DarkGreen] (0.25,3) arc(0:180:0.25);
    \draw[ultra thick, DarkGreen] (0.25,3)--(0.25,1);
     \draw[ultra thick, DarkGreen] (-0.25,3)--(-0.25,1);
  \draw[ultra thick, DarkGreen] (0.25,1) arc(180:360:0.25);
  \draw[ultra thick, DarkGreen] (0.75,3)--(0.75,1);
   \draw[ultra thick,DarkGreen] (1.25,3) arc(0:180:0.25);
  \draw[ultra thick, DarkGreen] (1.25,3)--(1.25,1);
    \draw [fill=Blue] (0,0) rectangle (0.4,0.1);
  \draw [fill=Red] (0.6,0) rectangle (1.0,0.1);
  \draw[]  (0,-0.2) rectangle (1.0,-0.1);
 \fill[left color = Blue,right color = Red]  (0,-0.2) rectangle (1.0,-0.1);
 \draw [fill=Black] (0,0.15) rectangle (0.2,0.22);
 \draw [fill=Black] (0.8,0.15) rectangle (1,0.22);
 \draw [fill=Black] (0.4,0.6) rectangle (0.6,0.67);
 \draw [fill=Black] (-0.33,1) rectangle (-0.4,1.2);
  \draw [fill=Black] (0.10,-1) rectangle (0.17,-1.2);
  \draw [fill=Black] (0.83,-1) rectangle (0.90,-1.2);
  \draw (0, 0.22)--(-0.4, 0.22);
  \draw (-0.4, 0.22)--(-0.4, 1);
  \draw (-0.2, 0.22)--(-0.2, -0.5);
  \draw (-0.2, -0.5)--(0.1, -0.5);
  \draw (0.1, -0.5)--(0.1, -1);
  \draw (1, 0.22)--(1.2, 0.22);
  \draw (1.2, 0.22)--(1.2, 0.6);
   \draw (1.2, 0.6)--(0.6, 0.6);
   \draw (1.2, 0.22)--(1.2, -0.5);
   \draw (1.2, -0.5)--(0.9, -0.5);
   \draw (0.9, -0.5)--(0.9, -1);
    \draw[ultra thick, Bittersweet] (-0.25,-1)--(-0.25,-3);
  \draw[ultra thick, Bittersweet] (-0.25,-3) arc(180:360:0.25);
    \draw[ultra thick, Bittersweet] (0.25,-1)--(0.25,-3);
     \draw[ultra thick,Bittersweet] (0.75,-1) arc(0:180:0.25);
  \draw[ultra thick, Bittersweet] (0.75,-1)--(0.75,-3);
   \draw[ultra thick, Bittersweet] (0.75,-3) arc(180:360:0.25);
  \draw[ultra thick, Bittersweet] (1.25,-1)--(1.25,-3);
   \node at (-0.7, 2) {$\hm{1}$};
        \node at (-0.7, -2) {$\hm{2}$};
     \end{tikzpicture}
     \caption{•}
                        \end{subfigure}
                                                               
                       \caption{$a)$ Possible circuit implementation of the sign flip of the coupling parameters with the transmons composing the TCQ coupled to two transmission line resonators of length $L$; $b)$ qualitative sketch of the circuit realization. Notice that we are using the same color coding as Fig. \ref{tcqLevelsTrans} and Fig. \ref{tcqLevelsChi120}.}
                       \label{figTCQ2res}
                         \end{figure}
In Fig. \ref{figTCQ2res} we see how building on this intuition we can conceive a configuration that implements the sign flip described by Eqs. \ref{signFlip}.The bare transmons composing the TCQ are coupled at the same location (or close to each other) to resonator $1$ while at locations separated by approximately half a wavelength to resonator $2$. This idea is somehow similar to what has been proposed in \cite{kockum2014} for implementing a giant atom coupled at different locations of a transmission line, although in a different context compared to the problem we are tackling here and most of all considering a simple transmon, and not a TCQ. \par 
To close this subsection, we point out that the circuit in Fig. \ref{figTCQ2res} implements only one TCQ. Thus, we need to add two further TCQs coupled in a similar way to both resonators in order to implement our parity measurement scheme.  
\section{Information Gain in dispersive parity measurement}
\label{sec5}
We now turn our attention to the problem of how to charachterize our parity measurement scheme in terms of information gains. Our approach will basically expand from the one in \cite{clerkNoise} and \cite{girvinNotes}, in which this problem was treated for the single qubit dispersive measurement. In our case, based on an information theoretical approach, we define what it is meant by information gain about a specific observable of the system. 
\par 
We consider the ideal case in which there is neither relaxation nor dephasing of the qubits. Like we assumed so far, the dispersive shifts $\chi_1$ and $\chi_2$, and the qubit-state dependent coupling of the resonators $\chi_{12}$, are matched, meaning that they are equal for all qubits. Within these assumptions the time evolution of the resonators' degrees of freedom depends only on the Hamming weight of the three qubits, and consequently also the output signal shows the same dependence.  The Hamming weight can assume values $h_w=\{0,1,2,3\}$. The information about the state of the qubits is encoded in the integrated signal of homodyne detection, which we call $I$. In particular, in the assumption of strong local oscillator this signal turns out to be Gaussian distributed with mean depending on the particular Hamming weight, which we generally write as

 \begin{equation}
\label{intSig}
I_{h_w}(\tau) = \int_{0}^{\tau} d t \biggl(\beta_{out, h_w}(t) e^{- i \phi}+\beta_{out, h_w}^*(t) e^{ i \phi}\biggr),
\end{equation}
with $\phi$ the phase of the local oscillator, and variance equal to $\tau$ (the measurement time) \footnote{For reference see in particular Chap.11, Eq. 11.4.30 in Ref. \cite{gardinerZollerNoise}.}. The Hamming weight dependent average of the output field $\beta_{out, h_w}(t)$ is obtained by averaging the input-output relation Eq. \ref{inputOutputRel}, assuming a specific initial Hamming weight of the qubits. Thus, the probability density for the random variable $I$ conditioned on a certain Hamming weight $h_w$, at a certain measurement time $\tau$ is
\begin{equation}
p(I | h_w)(\tau)= \frac{1}{\sqrt{2 \pi \tau}} \exp\biggl[-\frac{(I-I_{h_w}(\tau))^2}{2 \tau^2} \biggr].
\end{equation}
In order to define the information gain, let us consider the following scenario. Before the measurement we have complete ignorance about the state of the system. From our point of view the Hamming weight of the qubits can be whatever of the four possible values with equal probability $1/4$. Accordingly, we also assign uniform probability to the parity, which can be even or odd with probability $1/2$. At this point we measure and we ``learn" more about the state of the system via the observation of a certain realization of $I(\tau)$, which is our only information channel. Thus, the question we ask ourselves is ``what is the probability that a certain property of the system has a certain value, given the fact that we observed $I(\tau)$?". For instance, considering the Hamming weight we would like to obtain $p(h_w | I)$. We can obtain this conditional probability using Bayes theorem
\begin{equation}
p(h_w | I)= \frac{p(h_w) p(I | h_w)}{p(I)},
\end{equation}
where $p(I)$ is given by the sum of the probabilities of mutually exclusive events
\begin{equation}
p(I)= \sum_{i=0}^3 p(I |h_w=i) p(h_w=i).
\end{equation}
Notice that for notational simplicity we are omitting to write explicitly the dependency on the measurement time $\tau$.
Thus, since $p(h_w=i)=1/4 \, , \, \forall i$
\begin{equation}
p(h_w | I)= \frac{p(I | h_w)}{\sum_{i=0}^3 p(I |h_w=i)}.
\end{equation}
At this point, we define the information gain about the Hamming weight given a certain realization of the current $I$ as the difference between the final, conditional Shannon entropy and the initial one
\begin{multline}
\mathcal{I}_{h_w}(I)=\log_2 4+ \sum_{i=0}^3 p(h_w=i | I) \log_2 (p(h_w=i | I))= \\
2+\sum_{i=0}^3 p(h_w=i | I) \log_2 (p(h_w=i | I)).
\end{multline}
The average information gain about the Hamming weight is then defined as
\begin{equation}
\overline{\mathcal{I}}_{h_w}= \int_{-\infty}^{+\infty} dI p(I) \mathcal{I}_{h_w}(I).
\end{equation}
We can also define the Hamming weight measurement rate $\Gamma_{m, h_w}$ as the derivative of $\overline{\mathcal{I}}_{h_w}$ with respect to the measurement time $\tau$
\begin{equation}
\Gamma_{m, h_w} = \frac{d \overline{\mathcal{I}}_{h_w}}{d \tau}.
\end{equation}
$\overline{\mathcal{I}}_{h_w}$ can then be written as
\begin{equation}
\overline{\mathcal{I}}_{h_w}= \int_0^{\tau} dt \Gamma_{m, h_w} (t).
\end{equation}
According to our definition $\Gamma_{m, h_w}$ has the units of number of ``Hamming weight bits" per unit of time. Since, we need two bits in order to determine the Hamming weight the maximum measurement gain is $2$ bits. \par
We can actually repeat the previous discussion for any observable. In particular, we are interested in defining an average information gain about the parity $P$ of the three qubits. In this case we would be interested in the conditional probability that the parity $P$ is either even ($e$) or odd ($o$) given a certain realization of the current $I$ at a certain measurement time. These probabilities are given by
\begin{subequations}
\begin{multline}
p(P=e | I)= p (h_w=0 | I)+p(h_w=2 | I) = \\ \frac{1}{p(I)} [p(h_w=0)p(I | h_w=0)+p(h_w=2)p(I | h_w=2)],
\end{multline}
\begin{multline}
p(P=o | I)= p (h_w=1 | I)+p(h_w=3 | I) = \\ \frac{1}{p(I)} [p(h_w=1)p(I | h_w=1)+p(h_w=3)p(I | h_w=3)].
\end{multline}
\end{subequations}
We define the information gain about the parity given $I$ as the difference between the final conditional Shannon entropy and the initial one, \i.e., a uniform distribution with $p(P=e)=p(P=o)=1/2$
\begin{multline}
\mathcal{I}_P (I)= \log_2 2+\sum_{l=e,o} p(P=l | I) \log_2 p(P=l | I)= \\
1+\sum_{l=e,o} p(P=l | I) \log_2 p(P=l | I).
\end{multline}
The average information gain about the parity is
\begin{equation}
\overline{\mathcal{I}}_P = \int_{-\infty}^{+\infty} dI p(I) \mathcal{I}_{P}(I).
\end{equation}
We define a parity measurement rate $\Gamma_{m, P}$ as the derivative of the $\overline{\mathcal{I}}_P$ with respect to the measurement time
\begin{equation}
\Gamma_{m, P} = \frac{d \overline{\mathcal{I}}_{P}}{d \tau},
\end{equation}
which has units of "parity bits" per unit of time. The average parity information gain can then be written as
\begin{equation}
\overline{\mathcal{I}}_{P}= \int_0^{\tau} dt \Gamma_{m, P} (t).
\end{equation}
Since we need one bit in order to determine the parity of the three qubits the maximum parity information gain is $1$ bit. \par 
In order to compute these information gains, we have to solve the evolution equations for the field amplitudes of the two resonators depending on the Hamming weight of the three qubits. From these we can obtain the Hamming weight dependent output field amplitude and thus the integrated signal from Eq. \ref{intSig}. The evolution equations we have to solve are the following
\begin{subequations}
\label{eqalphahw}
\begin{multline}
\frac{d \alpha_{1, h_w}}{d t} = -i [\Delta_{d1}+\chi_1(3-2 h_w)] \alpha_{1, h_w}-\chi_{12}(3-2 h_w) \alpha_{2, h_w} \\ -\frac{\kappa_1}{2} \alpha_{1, h_w}-\frac{\sqrt{\kappa_1 \kappa_2}}{2} \alpha_{2, h_w} -i \sqrt{\kappa_1} \beta_{in}(t),
\end{multline}
\begin{multline}
\frac{d \alpha_{2, h_w}}{d t} = -i [\Delta_{d2}+\chi_2(3-2 h_w)] \alpha_{2, h_w}-\chi_{12}(3-2 h_w) \alpha_{1, h_w} \\ -\frac{\kappa_2}{2} \alpha_{2, h_w}-\frac{\sqrt{\kappa_1 \kappa_2}}{2} \alpha_{1, h_w} -i \sqrt{\kappa_2} \beta_{in}(t),
\end{multline}
\end{subequations}
from which we obtain the output field amplitude as $\beta_{out, h_w}=\beta_{in}(t)+ \sqrt{\kappa_1} \alpha_{1, h_w}+\sqrt{\kappa_2} \alpha_{2, h_w}$. $\beta_{in}(t)$ is the drive in this case, \i.e., the average of the input field $b_{in}$ comparing to Eq. \ref{inputOutputRel}.  The detuning between the resonator frequencies and the drive frequency, $\Delta_{d1}$ and $\Delta_{d2}$, are chosen like in Eqs. \ref{parityCondition} so that the output field depends only on the parity at the steady state. We will focus on the case in which $\kappa_1=\kappa_2= \kappa$, and we will report results using $\kappa$ as the unit of frequency. We also consider the case in which we are able to set $\chi_{12}=0$. As in \cite{criger2016}, we consider a realistic piecewise pulse
\begin{widetext}
\begin{equation}
\beta_{in}(t)= \begin{cases}
0 \quad &, \quad t< t_{on}, \\
\frac{\epsilon_{ss}}{2} \biggl[1-\cos \biggl(\frac{\pi}{\sigma}(t-t_{on}) \biggr) \biggr] \quad &, \quad t_{on} \le t < t_{on}+\sigma, \\
\epsilon_{ss} \quad &, \quad t_{on}+\sigma \le t < t_{off}, \\
\frac{\epsilon_{ss}}{2} \biggl[1+\cos \biggl(\frac{\pi}{\sigma}(t-t_{off}) \biggr) \biggr] \quad &, \quad t_{off}\le t < t_{off}+\sigma, \\
0 \quad &, \quad t \ge t_{off}+\sigma.
\end{cases}
\end{equation}
\end{widetext} 
We take a total measurement time $\tau= 28/\kappa$ \footnote{with a $\kappa/(2 \pi)= 5\, \mathrm{MHz}$ this corresponds to a measurement time of $0.70 \, \mu s$} and the following parameters of the pulse:
$\epsilon_{ss}= 0.5/\sqrt{\kappa}, \sigma= 4/28 \tau, t_{on}=1/28 \tau, t_{off}=16/28 \tau$. \par
\begin{figure}
\vspace{0.5cm}
\centering
\begin{subfigure}[t]{0.4 \textwidth}
\includegraphics[scale=0.5]{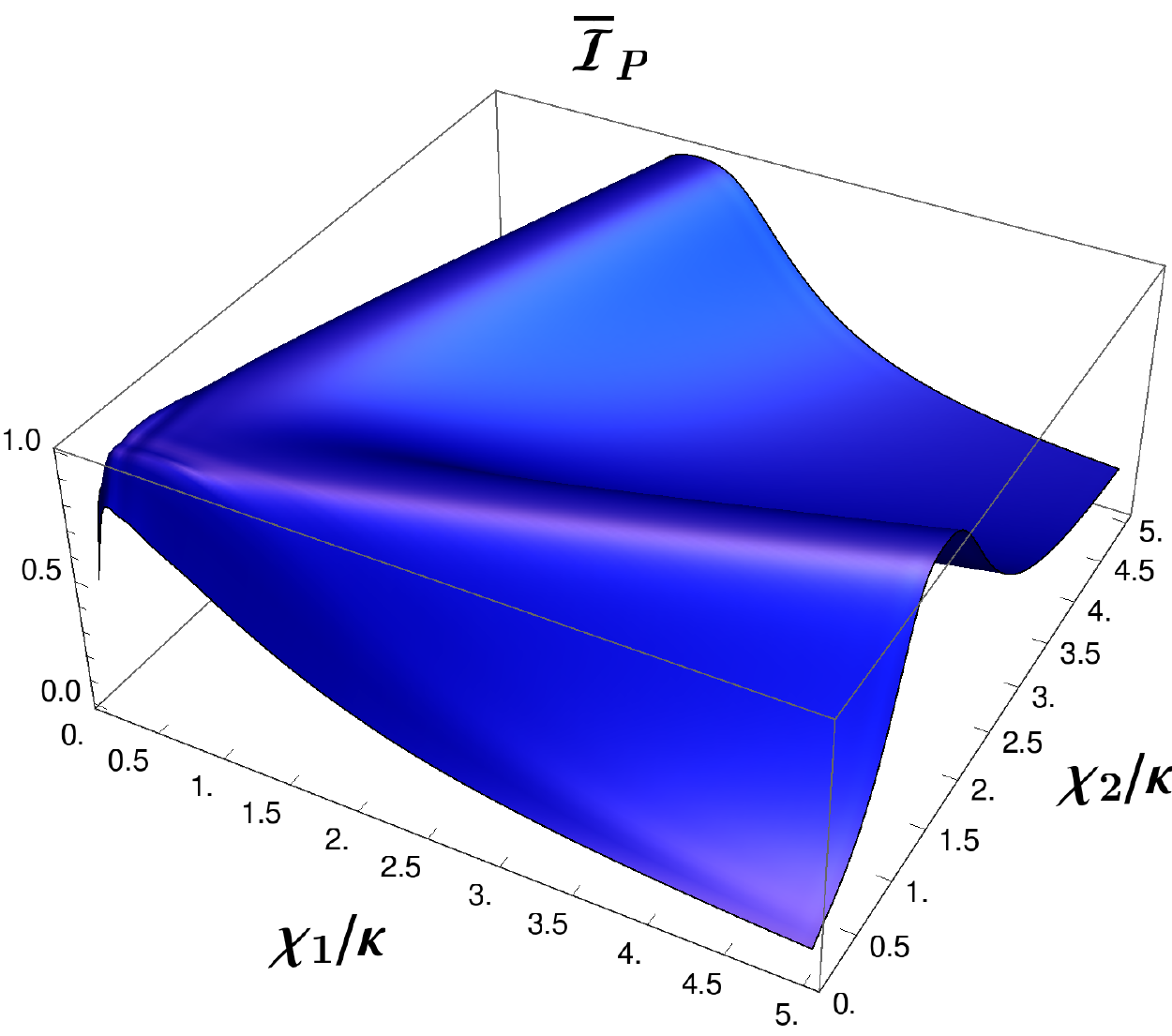}
\subcaption{}
\label{infGainFigA}
\end{subfigure}
\begin{subfigure}[t]{0.4 \textwidth}
\includegraphics[scale=0.6]{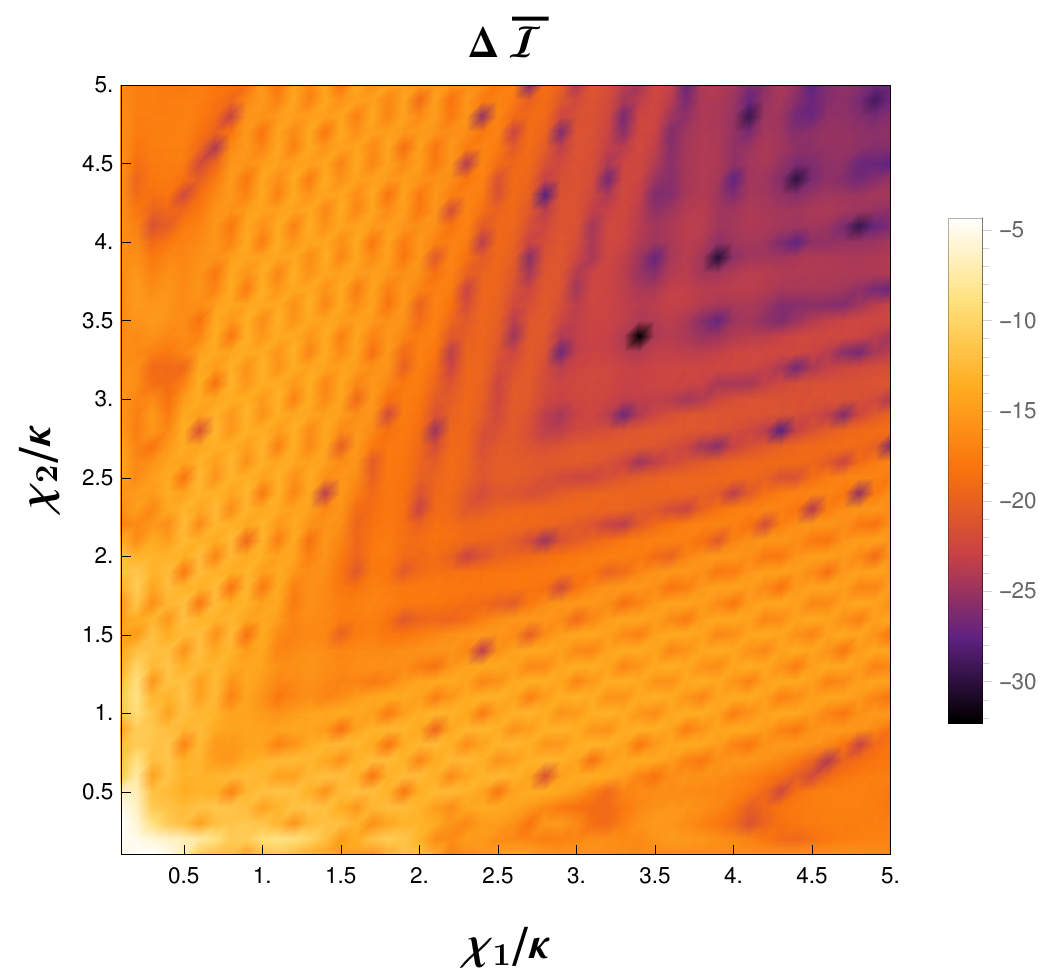}
\subcaption{}
\label{infGainFigB}
\end{subfigure}
\caption{$a)$ Average parity information gain for a measurement time $\tau \kappa= 28$ as a function of $\chi_1/\kappa$ and $\chi_2/\kappa$. The measured quadrature of the output field is always chosen to be the one that maximizes the information content about the state of the system. $b)$ Difference between information gain about the Hamming weight and parity on a $\log 10$ scale. We point out that the fine structure that can be noticed in this figure is an artifact of numerical interpolation.}
\label{infGainFig}
\end{figure}
\begin{figure}
\vspace{0.5cm}
\centering
\includegraphics[scale=0.5]{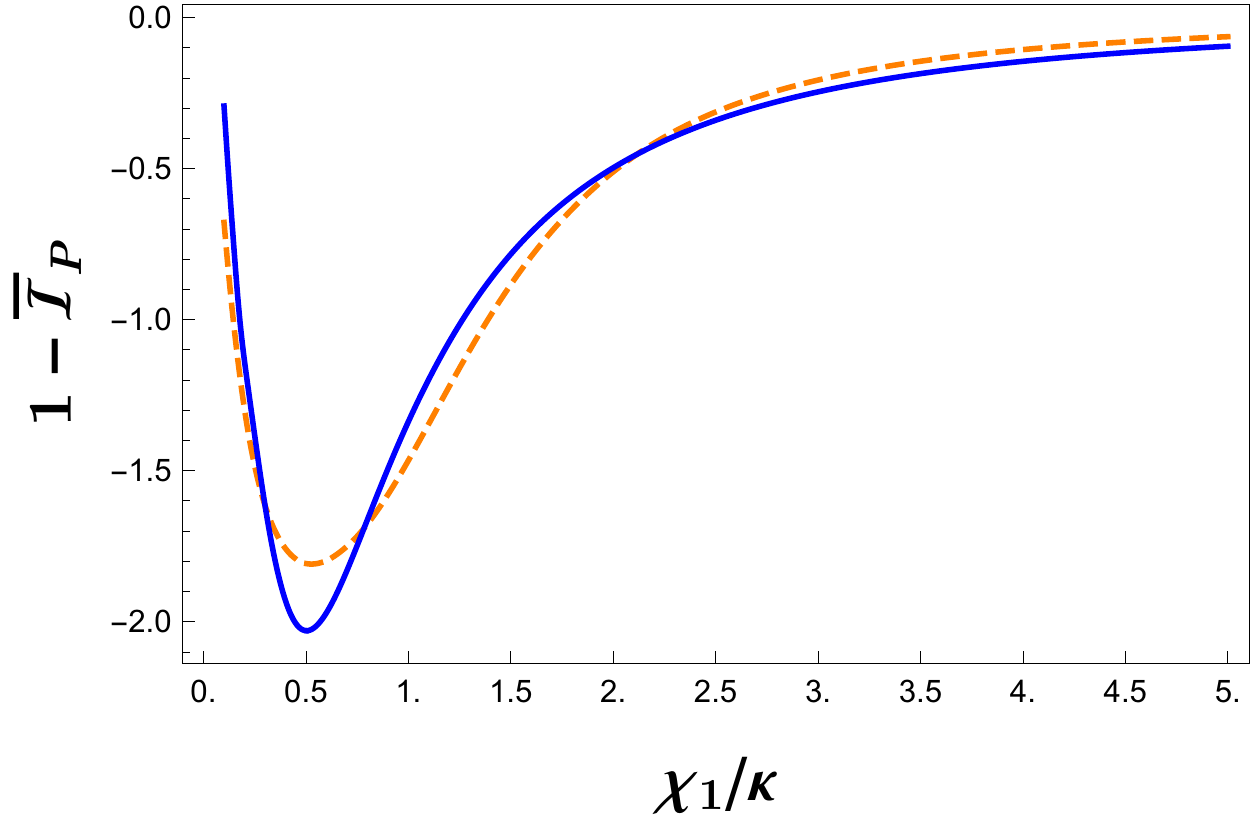}
\caption{Missing parity information on a $\log 10$ scale for the symmetric case $\chi_1=\chi_2$(blue solid line) and the asymmetric case in which $\chi_2=0.3 \kappa$ (orange dashed line). In both cases, we see a minimum of the missing information (maximum information gain) at $\chi_1 \approx \kappa/2$. The parameters are the same as in Fig. \ref{infGainFig}.}
\label{infGainFig2}
\end{figure}

In Fig. \ref{infGainFig} we clearly see that the measurement provides (to a good approximation) only information about the parity. Specifically, in Fig. \ref{infGainFigA} we see that there are regions of parameters in which we gained essentially complete information about the parity. In Fig. \ref{infGainFigB} we note that the difference between the $\mathcal{\overline{I}}_{h_w}$ and $\mathcal{\overline{I}}_{P}$, $\Delta \mathcal{\overline{I}}$ is always quite small, except in the region of small $\chi_1/\kappa$ and $\chi_2/\kappa$. This means that only the information about the parity of the Hamming weight is learned and not information about the additional bit. Analyzing Fig. \ref{infGainFigA} and \ref{infGainFigB} together, we realize that the optimal situation is the one in which we have approximately one bit of information about the parity, but very small difference between Hamming weight and parity information gain. In particular, if we consider the symmetric case $\chi_1=\chi_2=\chi$, we notice that we have maximum parity information gain at $\chi \approx \kappa/2$, which is the usual condition reported in the literature for maximizing the information gain in the standard single-qubit dispersive readout \cite{gambetta2008}. We see this more clearly in Fig. \ref{infGainFig2}, in which we plot the missing parity information as a function of $\chi_1/\kappa$ at the end of the measurement for the symmetric case and asymmetric case with $\chi_2=0.3 \kappa$. Even if the symmetric case gives a lower minimum, the asymmetric case might also be an interesting option when using a TCQ. In fact, if the TCQ is operated in the condition described in Subsec. \ref{achZerochi12}, it will experience a Purcell decay due only to resonator $2$. This means that we would like to have a small $\chi_2$, given by Eq. \ref{chi2TCQ}, in order to have a small $g_2$ and consequently a small Purcell effect.  
\subsection{Estimates of TCQ parameters}
Here we give some quantitative rough estimates of typical parameters that may be achievable using TCQs for our dispersive three-qubit parity measurement scheme. We consider directly the ideal configuration in which we are able to set $\chi_{12}=0$ for all TCQs. This means that we assume Eqs. \ref{signFlip} to hold for each TCQ. We denote the three TCQs by $\{a,b,c\}$. We take equal photon decay rates for the two resonators $\kappa_1=\kappa_2=\kappa$, with $\kappa /2 \pi= 5\, \mathrm{MHz}$ and set also $\chi_1 = \chi_2 = \chi= -\kappa /2$ \footnote{The minus sign change does not change the previous discussion.}. In this way, we achieve the optimal situation considered in the main part of this section, which is also the model studied in \cite{tornbergBarzanjeh, criger2016} via a stochastic master equation approach. The coupling parameter $J$ is also assumed for simplicity to be equal for all TCQs and in particular $J/2 \pi= -400 \, \mathrm{MHz}$. The energy difference between first excited state and ground state, \i.e., the levels that will form the qubit, are chosen as
$\tilde{\omega}_{-a}/2 \pi= 6000 \, \mathrm{MHz}$, $\tilde{\omega}_{-b}/2 \pi= 5600 \, \mathrm{MHz}$ and $\tilde{\omega}_{-c}/2 \pi= 5200 \, \mathrm{MHz}$. In this way, the qubits are sufficiently detuned so that we can neglect the resonator mediated interaction between them. The two transmons composing each TCQ are assumed to be almost resonant, so that, from Eq. \ref{lambda}, we can approximate for all TCQs $\lambda \approx \pi/4$ . Using this assumption $\tilde{\omega}_{+\alpha}=\tilde{\omega}_{-\alpha}- 2 J$ with $\alpha=\{a,b,c\}$. The anharmonicities are also taken to be all equal to $\delta/2 \pi=-E_c /2 \pi= -300 \, \mathrm{MHz}$. The frequency of the resonators are chosen $\omega_1/2 \pi= 7500 \, \mathrm{MHz}$ and $\omega_2/2 \pi= \omega_1+ 2 \sqrt{3} \chi$ in order to match the parity condition. With these parameters we obtain the following bare coupling parameters: $g_{1a}/2 \pi= 106.6 \, \mathrm{MHz}$, $g_{2 a}/2 \pi= 76.4  \,\mathrm{MHz}$, $g_{1b}/2 \pi= 132.5 \, \mathrm{MHz}$, $g_{2 b}/2 \pi= 113.3 \, \mathrm{MHz}$, $g_{1c}/2 \pi= 158.4  \, \mathrm{MHz}$, $g_{2 c}/2 \pi= 150.0\, \mathrm{MHz}$. We can also estimate the Purcell relaxation time for all TCQs as \cite{tcq2011} 
\begin{equation}
T_{p,\alpha}= \biggl[\kappa \biggl(\frac{\sqrt{2} g_{1,\alpha}}{\tilde{\omega}_{-\alpha}-\omega_1}\biggr)^2\biggr ]^{-1},
\end{equation}
 $\alpha=\{a,b,c\}$. In particular, we get $T_{p,a} \kappa= 100.1 $, $T_{p,b} \kappa= 103.7 $ and $T_{p,c} \kappa= 106.2 $. Comparing to the typical measurement time considered in the main part of the section $\tau \kappa= 28$ we see that the typical Purcell time is approximately between $3$ and $4$ times the measurement time for the symmetric case. As we mentioned in the main part of this section, longer Purcell times could be achieved in the condition $\chi_1=-\kappa/2$ and $\chi_2=-0.3 \kappa$. In this case, proceedeing as before we would get Purcell times $T_{p,a} \kappa= 166.8 $, $T_{p,b} \kappa=  172.8$ and $T_{p,c} \kappa= 177.0$. Notice, that from Fig. \ref{infGainFig2}, this case is charachterized by a higher missing information gain. Thus, we identify a tradeoff between information gain and Purcell relaxation time. However, the Purcell relaxation time may be additionally increased in both cases by means of Purcell filtering \cite{jeffrey2014, sete2015, bronn2015}.   

\section{Conclusions}
\label{Sec6}
Beyond the specific analysis of direct multi-qubit parity measurements, this paper provides a general analysis based on input-output theory for a system of two resonators coupled to the same transmission line; we expect these to be more applications for this general analysis.   
By identifying the general condition for obtaining a direct measurement of the parity of the string of qubits, we clarify the role of transient ring-up and ring-down signals in the collection of parity-only information; they are generally deleterious, but to a degree that depends on the exact choice of system parameters.  \par

Our formal analysis provides an example of the necessity of going beyond the RWA in some parts of the analysis, quite crucial in our case in properly assessing the role of the quantum switch term.  We feel that in the future, highly accurate modeling must go beyond the RWA in other ways, to assess a potential myriad of small effects; \cite{billangeon2015} provides an excellent example of developing such analysis in the context of a very different effective coupling scheme involving a multi-qubit, multi-resonator device. \par

Returning to our specific results, we have identified the main problem in implementing the direct multi-qubit parity scheme in the presence of qubit-state dependent coupling of the two resonators, emerging from the dispersive transformation of the Hamiltonian. We have shown that this problem can be solved by using a modified effective qubit, with the TCQ, a system of just double the complexity of an ordinary transmon, giving exactly the necessary greater flexibility to achieve the desired dispersive-coupling parameters. In particular, the harmful quantum switch terms is ideally cancelled using a TCQ as a qubit. To achieve this condition, it is essential that the sign of the bare Jaynes-Cummings coupling parameters be selective flipped.  This leads to the non-obvious strategy of coupling the transmons composing the TCQ to different locations of the transmission line resonators. \par

By uncovering the reason why this cancellation can happen, we are optimistic that the same reasoning may be applied to other systems beyond our TCQ-based approach. Our information-gain analysis of the parity measurement scheme, permitting us to quantify both the amount of extra parity information learned during the measurement as well as the typical time needed to complete the measurement, makes possible the determination of concrete optimal design parameters. The limitation due to Purcell relaxation is also straightforwardly assessed. \par

To conclude, we hope that this work will permit the community to objectively and systematically determine the best way forward towards fault tolerant, circuit-QED based quantum computing.  The parameter space is vast, so we expect that the kind of complete design analysis provided here will be essential for plotting our course into the future. 
\begin{acknowledgments}
We acknowledge financial support from the Alexander von Humboldt foundation and from the Excellence Initiative of the Deutsche Forschungsgemeinschaft.
\end{acknowledgments}
\appendix
\section{Dispersive Hamiltonian for the TCQ}
\label{appDispTCQ}
For the sake of completeness we report here the result for the dispersive Hamiltonian obtained for the first six levels of the TCQ: 
\begin{widetext}
\begin{multline}
\label{tcqDispTot}
\tilde{H}_{p d} = D \tilde{H}_p D^{\dagger} = H_0 + \sum_{i=1}^2\biggl \{ \frac{\tilde{g}_{i +}^2}{\tilde{\Delta}_{i +}} \ket{1_+ 0_-}\bra{1_+ 0_-}+\frac{\tilde{g}_{i -}^2}{\tilde{\Delta}_{i -}} \ket{0_+ 1_-}\bra{0_+ 1_-} +\frac{2 \tilde{g}_{i +}^2}{\tilde{\Delta}_{i+}+\tilde{\delta}_+} \ket{2_+ 0_-}\bra{2_+ 0_-} + \\
\frac{2 \tilde{g}_{i -}^2}{\tilde{\Delta}_{i-}+\tilde{\delta}_-} \ket{0_+ 2_-}\bra{0_+ 2_-}+ \biggl(\frac{\tilde{g}_{i +}^2}{\tilde{\Delta}_{i+}+\tilde{\delta}_c} +\frac{\tilde{g}_{i -}^2}{\tilde{\Delta}_{i-}+\tilde{\delta}_c}\biggr) \ket{1_+ 1_-} \bra	{1_+ 1_-} \biggr \}+ \\
 \sum_{i=1}^2 \biggl \{ \frac{\tilde{g}_{i+}^2}{\tilde{\Delta}_{i +}} \bigl(\ket{1_+ 0_-}\bra{1_+ 0_-}-\ket{0_+ 0_-}\bra{0_+ 0_-} \bigr)+ \frac{\tilde{g}_{i-}^2}{\tilde{\Delta}_{i -}} \bigl(\ket{0_+ 1_-}\bra{0_+ 1_-}-\ket{0_+ 0_-}\bra{0_+ 0_-} \bigr)+ \\ \frac{2 \tilde{g}_{i+}^2}{\tilde{\Delta}_{i +} +\tilde{\delta}_+} \bigl(\ket{2_+ 0_-}\bra{2_+ 0_-}-\ket{1_+ 0_-}\bra{1_+ 0_-} \bigr) + \frac{2 \tilde{g}_{i-}^2}{\tilde{\Delta}_{i -} +\tilde{\delta}_-} \bigl(\ket{0_+ 2_-}\bra{0_+ 2_-}-\ket{0_+ 1_-}\bra{0_+ 1_-}\bigr ) + \\ \frac{\tilde{g}_{i+}^2}{\tilde{\Delta}_{i +} +\tilde{\delta}_c} \bigl(\ket{1_+ 1_-}\bra{1_+ 1_-}-\ket{0_+ 1_-}\bra{0_+ 1_-}) + \frac{\tilde{g}_{i-}^2}{\tilde{\Delta}_{i -} +\tilde{\delta}_c} \bigl(\ket{1_+ 1_-}\bra{1_+ 1_-}-\ket{1_+ 0_-}\bra{1_+ 0_-}\bigr) \biggl \} a_i^{\dagger}a_i + \\
\biggl \{\frac{\tilde{g}_{1 +} \tilde{g}_{2+}}{2} \biggl(\frac{1}{\tilde{\Delta}_{1+}} +\frac{1}{\tilde{\Delta}_{2+}}\biggr) \bigl (\ket{1_+ 0_-}\bra{1_+ 0_-} - \ket{0_+ 0_-}\bra{0_+ 0_-} \bigr) +\frac{\tilde{g}_{1 -} \tilde{g}_{2-}}{2} \biggl(\frac{1}{\tilde{\Delta}_{1-}} +\frac{1}{\tilde{\Delta}_{2-}}\biggr) \bigl (\ket{0_+ 1_-}\bra{0_+ 1_-} - \ket{0_+ 0_-}\bra{0_+ 0_-} \bigr) + \\  \frac{2\tilde{g}_{1 +} \tilde{g}_{2+}}{2} \biggl(\frac{1}{\tilde{\Delta}_{1+}+\tilde{\delta}_{+}} +\frac{1}{\tilde{\Delta}_{2+}+\tilde{\delta}_+}\biggr) \bigl (\ket{2_+ 0_-}\bra{2_+ 0_-} - \ket{1_+ 0_-}\bra{1_+ 0_-} \bigr)+  \\ \frac{2\tilde{g}_{1 -} \tilde{g}_{2-}}{2} \biggl(\frac{1}{\tilde{\Delta}_{1-}+\tilde{\delta}_{-}} +\frac{1}{\tilde{\Delta}_{2-}+\tilde{\delta}_-}\biggr) \bigl (\ket{0_+ 2_-}\bra{0_+ 2_-} - \ket{0_+ 1_-}\bra{0_+ 1_-} \bigr) + \\ \frac{\tilde{g}_{1 +} \tilde{g}_{2+}}{2} \biggl(\frac{1}{\tilde{\Delta}_{1+}+\tilde{\delta}_{c}} +\frac{1}{\tilde{\Delta}_{2+}+\tilde{\delta}_c}\biggr) \bigl (\ket{1_+ 1_-}\bra{1_+ 1_-} - \ket{0_+ 1_-}\bra{0_+ 1_-} \bigr)+ \\ \frac{\tilde{g}_{1 -} \tilde{g}_{2-}}{2} \biggl(\frac{1}{\tilde{\Delta}_{1-}+\tilde{\delta}_{c}} +\frac{1}{\tilde{\Delta}_{2-}+\tilde{\delta}_c}\biggr) \bigl (\ket{1_+ 1_-}\bra{1_+ 1_-} - \ket{1_+ 0_-}\bra{1_+ 0_-} \bigr)   \biggr \} \bigl (a_1 a_2^{\dagger}+\mathrm{H.c.}\bigr)
\end{multline}
\end{widetext}
where for simplicity we directly omitted the two-photon transition terms, which can be neglected with the same argument as for the transmon.
\section{Transmon coupled to a transmission line resonator}
\label{appTransmonTL}
In this appendix we study a system made up of a transmon \cite{koch2007} capacitively coupled to a transmission line resonator of length $L$. Our goal is to obtain the generalized Jaynes-Cummings parameter as a function of the position of the transmon. This transmon can be one of those composing a TCQ for instance, and as we analyzed in Subsec. \ref{tcq2res}, we need to understand how the sign of the Jaynes-Cummings parameter of the bare transmons composing the TCQ behaves. A similar analysis is carried out in \cite{lalumiereInputOutput}. \\
Let us consider a transmission line resonator of length $L$ with a transmon capacitively coupled at a certain position $x_j$. In order to obtain the Lagrangian and then the Hamiltonian of this circuit it is appropriate to consider the discrete version of the circuit and afterwards take the continuous limit. The discrete circuit is shown in Fig. \ref{TLresTransmon}. Following standard references on circuit quantization \cite{devoret1997, bkd}, and taking the continuous limit introducing the flux field $\phi(x,t)$, the Lagrangian reads
\begin{multline}
\vspace{1cm}
\mathcal{L}=  \int_{0}^{L} dx \biggl \{  \frac{c}{2}\biggl(\frac{\partial \phi}{\partial t}  \biggr)^2 -\frac{1}{2 \ell}\biggl(\frac{\partial \phi}{\partial x} \biggr)^2 \biggr \}+\\
\frac{C_J}{2} \dot{\phi}_J^2 +E_J \cos \biggl(\frac{2 \pi \phi_J}{\phi_0} \biggr)+ \\
\int_{0}^{L} dx \frac{C_c \delta(x-x_J)}{2} \biggl(\frac{\partial \phi}{\partial t}- \dot{\phi}_J \biggr)^2.
\end{multline}
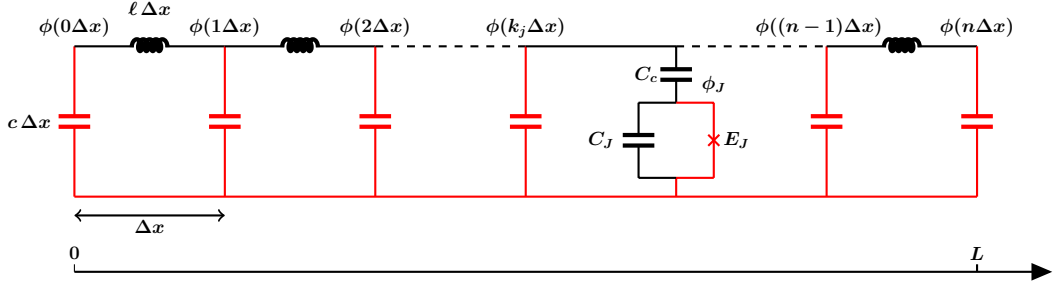
\begin{figure*}
\centering
 \begin{circuitikz}[american voltages, scale=0.5, every node/.style={transform shape}]
\draw [thick, -triangle 45] (-12, -1)--(14, -1);
\draw[thick, red] (-12,1)
to [short] (12,1);
\draw[thick, red] (-12,1)
to [C, color= red] (-12,5);
\draw[thick] (-12,5)
to [L] (-8,5);
\draw[thick, red] (-8,5)
to [C, color= red] (-8,1);
\draw[thick] (-8,5)
to [L] (-4,5);
\draw[thick, red] (-4,5)
to [C, color= red] (-4,1);
\draw [thick, dashed] (-4,5)
to (0,5);
\draw[thick, red] (0,5)
to [C, color=red] (0,1);
\draw[thick] (0,5)
to [short] (4,5)
to [C] (4,3.5); 
\draw[thick, red] (4,1)
to [short] (4,1.5)
to [short] (5, 1.5)
to [short] (5, 3.5)
to [short] (4, 3.5);
\draw[thick, red] (5, 2.5) node[cross=5,color=red,rotate=0] {};
\draw[thick] (4, 1.5)
to [short] (3, 1.5)
to [C] (3, 3.5)
to [short] (4, 3.5);
\draw[thick, dashed] (4, 5)
to [short] (8,5);
\draw [thick, red] (8,5)
to [C, color=red] (8,1);
\draw [thick] (8,5)
to [L] (12,5);
\draw [thick, red] (12,5)
to [C, color=red] (12,1);
\draw [thick, <->] (-12,0.5)
to (-8,0.5);
\node at (-13.2, 3) {\Large{$\hm{c\, \Delta x}$}};
\node at (-10, 0.2) {\Large{$\hm{\Delta x}$}};
\node at (-10,6) {\Large{$\hm{\ell \, \Delta x}$}};
\node at (-12, 5.5) {\Large{$\hm{\phi(0 \Delta x)}$}};
\node at (-8, 5.5) {\Large{$\hm{\phi(1 \Delta x)}$}};
\node at (-4, 5.5) {\Large{$\hm{\phi(2 \Delta x)}$}};
\node at (0, 5.5) {\Large{$\hm{\phi(k_j \Delta x)}$}};
\node at (5.6, 2.5) {\Large{$\hm{E_J}$}};
\node at (2, 2.5) {\Large{$\hm{C_J}$}};
\node at (3.2, 4.25) {\Large{$\hm{C_c}$}};
\node at (5, 4) {\Large{$\hm{\phi_J}$}};
\node at (7.8, 5.5) {\Large{$\hm{\phi((n-1) \Delta x)}$}};
\node at (12, 5.5) {\Large{$\hm{\phi(n \Delta x)}$}};
\draw (-12,-1) to (-12, -0.8);
\node at (-12, -0.5) {\Large{$\hm{0}$}};
\draw [thick] (12,-1) to (12, -0.8);
\node at (12, -0.5) {\Large{$\hm{L}$}};
 \end{circuitikz}
 \caption{Discrete circuit of a transmon capacitively coupled to a transmission line resonator. The spanning tree is depicted in red. The transmon is coupled capacitively at a certain position $k_j \Delta x$ with $k_j={1,2, \dots, n-2, n-1}$} and $\Delta x= L/n$. In the limiting procedure we will assume that the transmon is coupled at the desired position, \i.e., for $n \rightarrow +\infty$, $k_j \Delta x \rightarrow x_j$.
 \label{TLresTransmon}
\end{figure*}
Assuming that the coupling capacitance is smaller than the total capacitance of the line $c L$, which is a first weak coupling approximation, we can neglect its effect on the transmission line part of the Lagrangian and write
\begin{multline}
\label{lagrTLtransmon}
\mathcal{L}=  \int_{0}^{L} dx \biggl \{  \frac{c}{2}\biggl(\frac{\partial \phi}{\partial t}  \biggr)^2 -\frac{1}{2 \ell}\biggl(\frac{\partial \phi}{\partial x} \biggr)^2 \biggr \}+\\
\frac{C_{\Sigma}}{2} \dot{\phi}_J^2 +E_J \cos \biggl(\frac{2 \pi \phi_J}{\phi_0} \biggr)- \\
C_c \dot{\phi}_J \frac{\partial \phi}{\partial t} \Bigr|_{x=x_J},
\end{multline}
with $C_{\Sigma}= C_J+C_c$. Within this weak coupling assumption, we can obtain the normal modes of the transmission line; we will take open circuit boundary conditions \cite{girvinNotes}. We can write the flux $\phi(x,t)$ in terms of normal modes as
\begin{equation}
\phi(x,t)= \sum_{n=0}^{+\infty} \sqrt{2} \cos \biggl(\frac{\pi n}{L} x \biggr) \phi_n(t).
\end{equation} 
Introducing this expansion in the Lagrangian Eq. \ref{lagrTLtransmon}, we get
\begin{multline}
\label{lagrMode}
\mathcal{L}= \frac{1}{2}L c \biggl\{ \sum_{n=0}^{+\infty}\dot{\phi}_n^2-\omega_n^2 \phi_n^2 \biggr\}+ \\ 
\frac{C_{\Sigma}}{2} \dot{\phi}_J^2 +E_J \cos \biggl(\frac{2 \pi \phi_J}{\phi_0} \biggr)- \\
\dot{\phi}_J \sum_{n=0}^{n_c} C_{n}\dot{\phi}_n,
\end{multline}
where the mode frequencies are given by $\omega_n = \pi n/(\sqrt{\ell c} L)$ and we defined the coupling capacitances to the $n$-th mode 
\begin{equation}
C_{n}= C_{c} \sqrt{2} \cos \biggl(\frac{\pi n}{L} x_J \biggr).
\end{equation} 
In Eq. \ref{lagrMode}, we have introduced a phenomenological cutoff number $n_c \in \mathbb{N}$, and thus, an associated cutoff frequency $\omega_c$. This means that the modes with frequency higher than $\omega_c$ do not couple to the transmon. Although our original model does not give this cutoff, it is actually physically motivated. In fact, we have assumed that the coupling capacitance is a lumped element, \i.e., the capacitance per unit of length is proportional to a delta function. This is actually never the case and in general we should treat also the coupling capacitance as a distributed element. This means that when we consider modes whose associated wavelength is smaller than the typical length of the coupling capacitance, the quickly oscillating cosine functions tend to average out the effective coupling capacitance to the mode, and consequently decouple the mode from the transmon. Here, in order to keep the discussion simple, we just introduced a cutoff frequency, above which the modes do not couple to the transmon. \par  
In order to obtain the Hamiltonian, we first introduce the conjugate variables of the mode fluxes $\phi_n$ and the transmon flux $\phi_J$
\begin{subequations}
\label{conjVar}
\begin{equation}
q_n= \frac{\partial \mathcal{L}}{\partial \dot{\phi}_n}= L c \dot{\phi}_n -C_{n} \dot{\phi}_J [1-\Theta(n-n_c)],
\end{equation}
\begin{equation}
q_J= \frac{\partial \mathcal{L}}{\partial \dot{\phi}_J} = -\sum_{n=0}^{n_c} C_n \dot{\phi}_n + C_{\Sigma} \dot{\phi}_J,
\end{equation}
\end{subequations}
with $\Theta(x)$ the Heaviside step function, which we here define to have value $1$ at $x=0$.
The Hamiltonian is then obtained as the Legendre transform of the Lagrangian
\begin{equation}
\label{hamiltTLtransmon}
\mathcal{H}=q_J \dot{\phi}_J +\sum_{n=0}^{+\infty} q_n \dot{\phi}_n-\mathcal{L}.
\end{equation}
In order to write the Hamiltonian in terms of the conjugate variables we have to express the derivatives of the fluxes in terms of the conjugate variables. In particular, we have trivially 
\begin{equation}
\dot{\phi}_n =  \frac{q_n}{L c} \quad, \quad n> n_c.
\end{equation} 
For $n \le n_c$, we can write Eqs. \ref{conjVar} in matrix form as
\begin{equation}
\begin{bmatrix}
q_0 \\
q_1 \\
\vdots \\ 
q_{n_c} \\
q_J
\end{bmatrix} =\mathbf{C}
 \begin{bmatrix}
\dot{\phi}_0 \\
\dot{\phi}_1 \\
\vdots \\
\dot{\phi}_{n_c} \\
\dot{\phi}_J
\end{bmatrix},
\end{equation}
where we defined the capacitance matrix
\begin{equation}
\mathbf{C}= \begin{bmatrix}
L c & 0 & \dots & & -C_0 \\
0 & \ddots &  & & -C_1 \\
\vdots & &   & & \vdots \\
 & & &  &  -C_{n_c}\\
 -C_0 & -C_1 & \dots & -C_{n_c} & C_{\Sigma}.
\end{bmatrix}
\end{equation}
The problem is basically translated into the inversion of this matrix. We can carry out this matrix inversion via Block Matrix Inversion. In particular, we write our matrix in the following way
\begin{equation}
\mathbf{C}= \begin{bmatrix} \mathbf{A} & \mathbf{B} \\
\mathbf{B}^T & \mathbf{D}
\end{bmatrix},
\end{equation} 
where $\mathbf{A}$ is the square matrix
\begin{equation}
\mathbf{A}= L c \, \mathbf{I}_{n_c+1},
\end{equation}
with $\mathbf{I}_{n_c+1}$ the $(n_c+1)\times (n_c+1)$ identity matrix; $\mathbf{B}$ is the vector
\begin{equation}
\mathbf{B}= \begin{bmatrix}
-C_0 \\
-C_1 \\
\vdots \\
-C_{n_c} 
\end{bmatrix},
\vspace{2 mm}
\end{equation}
and $\mathbf{B}^{T}$ its transpose; finally $\mathbf{D}$ is the scalar
\begin{equation}
\mathbf{D}= C_{\Sigma}.
\end{equation}
The inverse of the capacitance matrix can be obtained as
\begin{widetext}
\begin{equation}
\mathbf{C}^{-1} = \begin{bmatrix}
\mathbf{A}^{-1}+\mathbf{A}^{-1} \mathbf{B} (\mathbf{D}-\mathbf{B}^T \mathbf{A}^{-1} \mathbf{B})^{-1} \mathbf{B}^T \mathbf{A}^{-1} & -\mathbf{A}^{-1} \mathbf{B} (\mathbf{D}-\mathbf{B}^T \mathbf{A}^{-1} \mathbf{B})^{-1} \\
-(\mathbf{D}-\mathbf{B}^T \mathbf{A}^{-1} \mathbf{B})^{-1} \mathbf{B}^T \mathbf{A}^{-1} & (\mathbf{D}-\mathbf{B}^T \mathbf{A}^{-1} \mathbf{B})^{-1} 
\end{bmatrix},
\end{equation} 
\end{widetext}
from which we get
\begin{equation}
\label{invC}
\mathbf{C}^{-1}= \begin{bmatrix}
\frac{1}{L c}(1+\frac{C_0^2}{\Sigma}) &  \frac{1}{L c} \frac{C_0 C_1}{\Sigma} & \dots & & \frac{C_0}{\Sigma} \\
\frac{1}{L c} \frac{C_0 C_1}{\Sigma} & \ddots &  & & \frac{C_1}{\Sigma} \\
\vdots & & & & \vdots \\
 & & &  & \frac{C_{n_c}}{\Sigma} \\
\frac{C_0}{\Sigma} & \frac{C_1}{\Sigma} & \dots & \frac{C_{n_c}}{\Sigma}& \frac{L c}{\Sigma}
\end{bmatrix},
\end{equation}
where we define
\begin{equation}
\label{sigma}
\Sigma= L c\, C_{\Sigma} -\sum_{n=0}^{n_c} C_n^2.
\end{equation}
Notice that without introducing the cutoff frequency the sum in Eq. \ref{sigma} would be a non-convergent series, although convergence would still be guaranteed in the more realistic model of a distributed coupling capacitance per unit of length. The problem of divergences in circuit QED has recently been tackled in \cite{gely2017}.\par 
At this point, we make again a weak coupling assumption approximating $\Sigma \approx Lc \, C_{\Sigma}$ and neglecting all terms $C_k C_l/(Lc \, C_{\Sigma})$ in the inverse of the coupling capacitance. Thus, we approximate
\begin{equation}
\mathbf{C}^{-1} \approx \begin{bmatrix}
\frac{1}{L c} &  0 & \dots & & \frac{C_0}{L c \, C_{\Sigma}} \\
0 & \ddots &  & & \frac{C_1}{L c \, C_{\Sigma}} \\
\vdots & & & & \vdots \\
 & & &  &  \frac{C_{n_c}}{L c \, C_{\Sigma}}\\
\frac{C_0}{L c \, C_{\Sigma}} & \frac{C_1}{L c \, C_{\Sigma}} & \dots & \frac{C_{n_c}}{L c \, C_{\Sigma}} & \frac{1}{ C_{\Sigma}},
\end{bmatrix}
\end{equation}
which basically means that we can write
\begin{subequations}
\label{phiFq}
\begin{equation}
\dot{\phi}_n= \frac{1}{Lc} q_n + \frac{C_n}{L c C_{\Sigma}} q_J \, , \, n=\{0,1,\dots, n_c\},
\end{equation}
\vspace{1mm}
\begin{equation}
\dot{\phi}_J= \sum_{n=0}^{n_c}\frac{C_n}{L c C_{\Sigma}} q_n+\frac{1}{C_{\Sigma}} q_{J}.
\end{equation}
\end{subequations}
Substituting Eqs. \ref{phiFq} into the Hamiltonian Eq. \ref{hamiltTLtransmon}, and again neglecting terms $C_k C_l/(Lc \, C_{\Sigma})$, we finally obtain the Hamiltonian
\begin{multline}
\mathcal{H}= \sum_{n=0}^{+\infty} \frac{q_n^2}{2 L c} + \frac{1}{2} L c \omega_n^2 \phi_n^2 +\\ \frac{q_J^2}{2 C_{\Sigma}} - E_J \cos \biggl(\frac{2 \pi \phi_J}{\phi_0} \biggr) + \\
\sum_{n=0}^{n_c} \frac{C_n}{L c C_{\Sigma}} q_j q_n.
\end{multline}
We point out that so far everything is classical. We now promote the flux variables and the related conjugate variables to operators imposing the usual commutation relations. In addition, we introduce annihilation and creation operators for the harmonic oscillator modes, \i.e., $n \ge 1$ as 
\begin{subequations}
\begin{equation}
\phi_n= \sqrt{\frac{\hbar }{2 L\, c \, \omega_n}}\bigl(a_n+a_n^{\dagger}\bigr),
\end{equation}
\begin{equation}
q_n = i \sqrt{\frac{\hbar L\, c \, \omega_n}{2}}\bigl(a_n^{\dagger}-a_n \bigr).
\end{equation}
\end{subequations}
and approximating the transmon as a Duffing oscillator, with Hamiltonian given by Eq. \ref{HtransmonRes}, we also introduce annihilation and creation operators for the transmon \cite{koch2007},
\begin{subequations}
\begin{equation}
\phi_J= \frac{\hbar}{2 e} \biggl(\frac{2 E_C}{E_J} \biggr)^{1/4}\bigl(b+b^{\dagger}\bigr),
\end{equation}
\begin{equation}
q_J = i 2 e \biggl (\frac{E_J}{32 E_C}\biggr)^{1/4} \bigl(b^{\dagger}-b \bigr).
\end{equation}
\end{subequations}
We finally, write the quantum Hamiltonian as 
\begin{multline}
H= \frac{q_0^2}{2 L c} + \sum_{n=1}^{+\infty} \omega_n a_n^{\dagger} a_n +H_{duff} +  \frac{C_0}{(L c)^2} q_J q_0 - \\ 2 e \sum_{n=1}^{n_c} \frac{C_n}{C_{\Sigma}} V_{rms,n} \biggl( \frac{E_J}{32 E_C} \biggr)^{1/4}(b^{\dagger}-b)(a_n^{\dagger}-a_n),
\end{multline}
with $V_{rms,n}= \sqrt{\hbar \omega_n/(2 L c)}$. Performing a RWA, neglecting fast rotating terms $b^{\dagger}a_n^{\dagger}$ and $b a_n$, we finally end up with a generalized multi-mode Jaynes-Cummings model with interaction Hamiltonian 
\begin{equation}
H_{int}= +2 e \sum_{n=1}^{n_c} \frac{C_n}{C_{\Sigma}} V_{rms,n} \biggl( \frac{E_J}{32 E_C} \biggr)^{1/4}(b^{\dagger}a_n+b a_n^{\dagger}),
\end{equation}
where we identify the generalized Jaynes-Cummings coupling parameter to mode $n$ as a function of the position $x_J$ of the transmon as 
\begin{equation}
\label{gxj}
g_n(x_J) = 2 e \frac{C_c}{C_{\Sigma}} V_{rms,n} \sqrt{2} \cos \biggl(\frac{\pi n}{L} x_{J} \biggr),
\end{equation}
with $1 \le n \le n_c$.

\bibliography{paperTCQbibliography}

\end{document}